\documentclass[12pt]{article}

\usepackage{amsmath,graphicx}
\usepackage{multirow}
\usepackage{amsfonts}
\usepackage{amssymb}
\usepackage{amscd}
\usepackage{cite}
\usepackage{amsmath}
\usepackage{bbm}

\def\hybrid{\topmargin -20pt    \oddsidemargin 0pt
        \headheight 0pt \headsep 0pt
        \textwidth 6.25in       
        \textheight 9.5in       
        \marginparwidth .875in
        \parskip 5pt plus 1pt   \jot = 1.5ex}

\hybrid

\numberwithin{equation}{section}
\numberwithin{table}{section}

\newcommand{\be}{\begin{equation}}
\newcommand{\ee}{\end{equation}}
\newcommand{\M}{{\bf M}}
\newcommand{\bea}{\begin{eqnarray}}
\newcommand{\eea}{\end{eqnarray}}

\def\cN{{\cal N}}
\def\cF{{\cal F}}
\def\cG{{\cal G}}
\def\nv{n_{\rm v}}
\def\nh{n_{\rm h}}
\renewcommand{\Re}{\operatorname{Re}}
\renewcommand{\Im}{\operatorname{Im}}
\newcommand\e{\mathrm{e}}
\newcommand\iu{\operatorname{i}}
\newcommand\diff{\mathrm{d}}

\newcommand{\pt}{\partial}

\def\a{\alpha}
\def\L{I}

\def\ax{{\tilde \phi}} 
\def\Kdil{{k_\phi}}    
\def\Kax{{k_{\tilde \phi}}}  
\def\Kxi{{k}}                
\def\Ktxi{{\tilde k}}        
\def\kk{{k}}          

\def\Proj{{\Pi}}

\def\P{{r}}  
\def\Q{{s}}  
\def\A{{t}}  
\def\R{{R}}  
\def\C{C}    
\def\D{D}    
\def\E{\rho}

\begin{document}
\begin{titlepage}
\begin{center}
\rightline{\small ZMP-HH/09-29}
\vskip 1cm

{\Large \bf
Spontaneous $\cN=2 \rightarrow \cN=1$ Supersymmetry Breaking in Supergravity and Type II String Theory}	
\vskip 1.2cm
{\bf Jan Louis$^{a,b}$, Paul Smyth$^{a}$, and Hagen Triendl$^{a}$,}

\vskip 0.8cm

$^{a}${\em II. Institut f\"ur Theoretische Physik der Universit\"at Hamburg, Luruper Chaussee 149, 22761 Hamburg, Germany}
\vskip 0.4cm

{}$^{b}${\em Zentrum f\"ur Mathematische Physik,
Universit\"at Hamburg,\\
Bundesstrasse 55, D-20146 Hamburg}
\vskip 0.8cm

{\tt jan.louis,paul.smyth,hagen.triendl@desy.de}

\end{center}

\vskip 1cm

\begin{center} {\bf ABSTRACT } \end{center}

\noindent
Using the embedding tensor formalism we give the general conditions for the existence of $\cN=1$ vacua in spontaneously broken $\cN=2$ supergravities. Our results confirm the necessity of having both electrically and magnetically charged multiplets in the spectrum, but also show that no further constraints on the special K\"ahler geometry of the vector multiplets arise. The quaternionic field space of the hypermultiplets must have two commuting isometries, and as an example we discuss the special quaternionic-K\"ahler geometries which appear in the low-energy limit of type II string theories. For these cases we find the general solution for stable Minkowski and AdS $\cN=1$ vacua, and determine the charges in terms of the holomorphic prepotentials. We find that the string theory realisation of the $\cN=1$ Minkowski vacua requires the presence of non-geometric fluxes, whereas they are not needed for the AdS vacua. We also argue that our results should hold in the presence of spacetime and worldsheet instanton corrections.
\vfill

\end{titlepage}

\section{Introduction}
\label{section:susybreaking}
A supersymmetric theory usually has a Minkowski or anti-de Sitter (AdS) ground state which preserves all of its supercharges. It is also possible -- albeit more difficult -- to construct models where supersymmetry is spontaneously broken. In this case, if one insists on a Lorentz-invariant ground state, supersymmetry generically breaks completely leaving no unbroken supercharges intact. The formulation of this result as a no-go theorem for partial supersymmetry breaking in Minkowski space has been known for a long time \cite{Cecotti:1984rk,Cecotti:1984wn}. The possibility of partial supersymmetry breaking in globally $\cN=2$ supersymmetric theories in four spacetime dimensions ($D=4$) was subsequently discovered some time later \cite{Bagger:1994vj,Antoniadis:1995vb}. In particular, it was observed in \cite{Antoniadis:1995vb} that the presence of a magnetic Fayet-Iliopoulos term spontaneously breaks $\cN=2\to \cN=1$. The supergravity version of this situation was presented in \cite{Ferrara:1995gu,Ferrara:1995xi,Fre:1996js} for a specific class of gauged $\cN=2$ theories. There it was found that the no-go theorem of \cite{Cecotti:1984rk,Cecotti:1984wn} could be avoided in a specific basis for the scalar fields of the $\cN=2$ vector multiplets. Apart from a few explicit examples which do show the possibility of partial $\cN=2\to \cN=1$ breaking, a more systematic analysis of the problem has been missing so far (see \cite{Hohm:2004rc} for an analysis in $D=3$). Our goal is to close this gap, by finding and then solving the general conditions in $\cN=2$ supergravity for partial supersymmetry breaking in Minkowski and anti-de Sitter spacetimes.

The second motivation of this paper comes from string theory, where flux compactifications on generalised geometries have been much discussed \cite{Grana:2005jc,Douglas:2006es,Blumenhagen:2006ci,Wecht:2007wu,Samtleben:2008pe}. Generically the resulting low-energy effective theory is a gauged supergravity with a scalar potential which lifts (part of) the vacuum degeneracy. It is clearly of interest to determine the amount of supersymmetry preserved by the ground state and how many supercharges are spontaneously broken. However, in classical gravity one is faced with another no-go theorem, due to Gibbons \cite{Gibbons:1984kp}, de Wit et al.\ \cite{deWit:1986xg} and Maldacena and Nu\~{n}ez \cite{Maldacena:2000mw}, which forbids flux compactifications to Minkowski space in the absence of negative energy-density sources, regardless of the amount of supersymmetry preserved. Furthermore, in \cite{Mayr:2000hh} it was noted that even if one evades the various no-go theorems and finds an $\cN=1$ vacuum, worldsheet instanton corrections in $\cN=2$ flux compactifications could ruin the result and reinstate the no-go theorem forbidding partial supersymmetry breaking.

In the known examples of partial supersymmetry breaking, the standard holomorphic prepotential does not exist as one of the gauge bosons has been exchanged with its magnetic dual via a symplectic rotation \cite{Ceresole:1995jg}. The lack of a prepotential makes it difficult to form a general picture. Therefore, it is advantageous to reinstate the prepotential, which one can always do at the expense of having to introduce both electric and magnetic charges. It turns out that the recently developed embedding tensor formalism \cite{deWit:2002vt,deWit:2005ub} is ideally suited to address this problem. This formalism treats electric and magnetic gauge bosons on the same footing and the conditions for partial supersymmetry breaking can then be formulated as a condition on the embedding tensor itself.
We shall see that this condition can be solved for any moduli space that admits an appropriate pair of Killing vectors.
This allows us to construct a general solution for Minkowski and AdS vacua displaying  $\cN=1$ supersymmetry for a broad class of $\cN=2$ gauged supergravities. More precisely, we give the construction of embedding tensors that lead to $\cN=1$ vacua for any moduli space that admits the Heisenberg algebra of Killing vectors naturally appearing in flux compactifications of type II string theory. Moreover, we find that by adjusting the charges one can realise $\cN=1$ vacua at any point of the moduli space.

In the second part of this work we discuss the uplift of these gauged supergravities to flux compactifications. We show that by rewriting the conditions for partial supersymmetry breaking in terms of the embedding tensor, the compactification no-go theorem can be evaded by including non-geometric fluxes. As we are able to phrase the conditions for an $\cN=1$ vacuum in terms of a general holomorphic prepotential, this also opens the possibility of finding solutions in the presence of instanton corrections. Finally, the flux quantisation condition forces the embedding tensor to have integer entries only, leading to a lattice in the moduli space where $\cN=1$ vacua can be realised. More importantly, this might restrict the possibility of $\cN=1$ vacua to a subclass of moduli spaces.

This paper is organised as follows. In Section~\ref{section:N=2} we review ${\cal N}=2$  supergravity in four dimensions and discuss the formulation of the theory with electric and magnetic gaugings in terms of the embedding tensor. We pay particular attention to the symplectic extension of the supersymmetry variations and the scalar potential, which play an important role throughout. In Section~\ref{section:vectors} we recall the basic facts of partial supersymmetry breaking. Focussing on the gravitino and gaugino sector, we then rederive the ``2 into 1 won't go'' theorem for Minkowski vacua in \ref{section:no-go} and show how it can be evaded by including both electric and magnetic charges in \ref{section:magnetic_vectors}. This allows us to rewrite the necessary conditions for spontaneous partial supersymmetry breaking in the gravity plus vector multiplet sector in a compact form and to solve them for general moduli spaces. Subsequently we analyse the conditions arising from the hyperino variations in Section~\ref{section:hypermultiplets}, restricting ourselves to the isometries that naturally arise in flux compactifications. We then present the full set of conditions for spontaneous $\cN=2$ to $\cN=1$ supersymmetry breaking in Minkowski and AdS spacetimes. In Section~\ref{section:strings}, we discuss the relation of our results to flux compactifications of string theory and the effect of string and brane instanton corrections. We identify the type of flux required for partial supersymmetry breaking, finding that Minkowski vacua require non-geometric fluxes, whereas AdS vacua may be found with geometric fluxes alone. We present our conclusions in Section~\ref{section:conclusions}. A collection of useful formula from $\cN=2$ supergravity are presented in Appendix~\ref{appA}. In Appendix~\ref{section:stability} we prove the stability of the ${\cal N}=1$ vacua by directly considering the potential and its derivatives, and rederive the Breitenlohner-Freedman bound for the AdS case.

While we were completing this paper \cite{Cassani:2009na} appeared, which should have some overlap with our discussion of $\cN=1$ AdS vacua.


\section{Gauged ${\cal N}=2$ Supergravity with Electric and Magnetic Charges}
\label{section:N=2}

In $D=4$ the spectrum of ${\cal N}=2$ supergravity consists of a gravitational multiplet, $\nv$ vector multiplets and $\nh$hypermultiplets (see e.g.\ \cite{Andrianopoli:1996cm}).\footnote{There also is the possibility of having tensor multiplets in the spectrum. For the purpose of this paper we dualise them to hypermultiplets if they are massless and to vector multiplets if they are massive.} The gravitational multiplet $(g_{\mu\nu},\Psi_{\mu {\cal A}}, A_\mu^0)$ contains the spacetime metric $ g_{\mu\nu}, \mu,\nu =0,\ldots,3$, two gravitini $\Psi_{\mu {\cal A}}, {\cal A}=1,2$ and the graviphoton $A_\mu^0$. A vector multiplet $(A_\mu,\lambda^{\cal A}, t)$ contains a vector $A_\mu$, two gaugini $\lambda^{\cal A}$  and a complex scalar $t$. Finally, a hypermultiplet $(\zeta_{\alpha}, q^u)$ contains two hyperini $\zeta_{\alpha}$ and 4 real scalars $q^u$. For $\nv$ vector- and $\nh$ hypermultiplets there are a total of $2\nv +4\nh$ real scalar fields and $2(\nv+\nh)$ spin-$\tfrac12$ fermions in the spectrum. For an Abelian, ungauged theory the bosonic Lagrangian is given by
\begin{equation}\begin{aligned}\label{sigmaint}
{\cal L}\ =\  - \mathrm{Im} \mathcal{N}_{IJ}\,F^{I}_{\mu\nu}F^{\mu\nu\, J}
- \mathrm{Re} \mathcal{N}_{IJ}\,
F^{I}_{\mu\nu} F_{\rho\sigma}^{J}\epsilon^{\mu\nu\rho\sigma}
+ g_{i\bar \jmath}(t,\bar t)\, \partial_\mu t^i \partial^\mu\bar t^{\bar \jmath}
+ h_{uv}(q)\, \partial_\mu q^u \partial^\mu q^v
\ ,
\end{aligned}\end{equation}
where the field strengths $F^{I}_{\mu\nu}, I=0,\ldots, \nv$ include the
graviphoton and their kinetic matrix $
\mathcal{N}_{IJ}$ is a function of the $\nv$ scalars $t^i,\, i =
1,\ldots,\nv $.\footnote{$\mathcal{N}_{IJ}$ is given in terms of the holomorphic prepotential
in \eqref{Ndef}.}
The scalar field kinetic terms form a $\sigma$-model Lagrangian, where
$(t^i,q^u),\, u=1,\ldots,4\nh,$ are viewed as coordinates on the manifold
\begin{equation}\label{scalarM}
{\M} = {\M}_{\rm v}\times {\M}_{\rm h} \ .
\end{equation}

In the hypermultiplet sector $h_{uv}$ denotes the metric on the $4\nh$-dimensional space  ${\M}_{\rm h}$, which ${\cal N}=2$ supersymmetry constrains to be a quaternionic-K\"ahler manifold \cite{Bagger:1983tt,deWit:1984px}.  Such manifolds have a holonomy group given by $Sp(1)\times Sp(\nh)$. In addition, they admit a
triplet of complex structures $J^x, x=1,2,3$ which satisfy the quaternionic algebra $J^x J^y =
-\delta^{xy}{\bf 1} + \epsilon^{xyz} J^z$. The metric $h_{uv}$ is Hermitian with respect to all
three complex structures. Correspondingly, a  quaternionic-K\"ahler manifold admits a triplet of hyper-K\"ahler forms $K^x_{uv} = h_{uw} (J^x)^w_v$ which are only covariantly closed with respect to the $Sp(1)$ connection $\omega^x$, contrary to the case of hyper-K\"ahler manifolds,
i.e.\
\begin{equation}\label{deriv_Sp(1)_curvature}
\nabla K^x \equiv dK^x + \epsilon^{xyz} \omega^y \wedge K^z=0 \ .
\end{equation}
In other words, $K^x$ is proportional to the $Sp(1)$ field strength of $\omega^x$, thus leading to
\begin{equation} \label{def_Sp(1)_curvature}
 K^x =\diff \omega^x + \tfrac12 \epsilon^{xyz} \omega^y\wedge \omega^z\ .
\end{equation}

In the vector multiplet sector $g_{i\bar \jmath}$ denotes the metric
of the $2\nv$-dimensional space ${\M}_{\rm v}$, which
${\cal N}=2$ supersymmetry constrains to be a  special
K\"ahler manifold \cite{deWit:1984pk,Craps:1997gp}. This implies that
the metric obeys
\begin{equation}\label{Kdef}
g_{i\bar \jmath} = \partial_i \partial_{\bar \jmath} K^{\rm v}\ ,
\qquad \textrm{for}\qquad
K^{\rm v}= -\ln \iu \left( \bar X^I \cF_I - X^I\bar \cF_I \right)\ .
\end{equation}
Both $X^I(t)$ and ${\cal F}_I(t)$ are holomorphic functions of the scalars
$t^i$ and in the ungauged case one can always choose
$\cF_I = \partial\cF/\partial{X^I}$, i.e.\ $\cF_I$
is the derivative of a holomorphic prepotential $\cF(X)$  which
is homogeneous of degree two. Furthermore, using coordinate invariance
it is possible to go to a system of `special coordinates' where
$X^I= (1,t^i)$. For more details on the vector multiplet sector in $\cN=2$ supergravity, see Appendix~\ref{section:sg}.

The equations of motion derived from $\cal L$  are invariant under $Sp(\nv+1)$ electric-magnetic duality rotations which act on the $(2\nv+2)$-dimensional symplectic vectors $(F^I, G_I)$ and $(X^I, \cF_I)$. $G_I$ are the field strengths of the dual magnetic vectors which only appear on-shell, in that they are not part of the Lagrangian \eqref{sigmaint}. Due to the symplectic invariance it is a matter of convention which vector fields are called electric and which are called magnetic. It is customary to denote the gauge fields which do appear in $\cal L$ as electric and their duals as magnetic.

The situation is more complicated in the presence of charged scalars, i.e.\ in gauged supergravities \cite{Louis:2002ny,deWit:2002vt,Dall'Agata:2003yr,Sommovigo:2004vj,D'Auria:2004yi,deWit:2005ub}. The charges appearing in the gauging break the symplectic invariance and the resulting theory crucially depends on which charges (electric or magnetic) the fermions and scalars carry. If all matter fields carry only electric charges, i.e.\ are charged with respect to the gauge fields which are declared electric in the ungauged case, then the Lagrangian is given by a standard $\cN=2$ gauged supergravity. However, it is possible that some fraction of the matter fields also carry magnetic charges, as frequently occurs in string compactifications. In this case it is still possible to symplectically rotate the vectors to the electric frame, such that all the matter fields are electrically charged i.e.\ the initial electric and  magnetic charges are constrained to be mutually local. However, as the theory is no longer symplectically covariant the Lagrangian in the electric frame might not be of the standard supergravity form . In particular, $\cF_I$ is no longer constrained to be the derivative of a prepotential, or in other words, a holomorphic $\cal F$ might not exist in the given symplectic frame  \cite{Ceresole:1995jg,deWit:2002vt,deWit:2005ub}.

As we shall see, one of the necessary conditions for partial supersymmetry breaking is precisely the existence of magnetically charged fields. Therefore, the formalism of the embedding tensor introduced in \cite{deWit:2002vt,deWit:2005ub} is ideally suited to discuss the problem of partial supersymmetry breaking. It treats the electric vectors $A_\mu^{~~\L}$ and the magnetic vectors $B_{\mu\L}$ on the same footing and naturally allows for arbitrary gaugings. Let us now briefly introduce the formalism, following \cite{deWit:2002vt,deWit:2005ub}.

The $\cN=2$ theory has a group $G_0$ of global isometries on $\M$ generated by the Killing vectors $k_{\hat\a}, \hat \a = 1,\ldots, {\rm dim}(G_0)$. The embedding tensor $ \Theta_\Lambda^{~~\hat\a}$ has electric and magnetic components, which we denote as $ \Theta_\Lambda^{~~\hat\a} = (\Theta_\L^{~~\hat\a},-\Theta^{\L\hat\a})$, where $\Lambda$ labels the $(2\nv+2)$ electric and magnetic gauge fields.\footnote{Here the minus sign in $\Theta_\Lambda^{~~\hat\a}$ is introduced such that $\Theta^{\Lambda\hat\a} = \Omega^{\Lambda \Sigma}  \Theta_{\Lambda}^{~~\hat\a}= (\Theta^{\L\hat\a}, \Theta_\L^{~~\hat\a})$ transforms covariantly under symplectic rotations, where $\Omega^{\Lambda\Sigma}$ is the inverse $Sp(\nv+1)$ metric, cf.\ Appendix~\ref{section:sg}.} 
Given the set of global generators $k_{\hat\a}$, the embedding tensor selects the gauged subset
\begin{equation}\label{generators}
Y_\Lambda =
\Theta_\Lambda^{~~\hat\a} k_{\hat\a} = (\Theta_\L^{~~\hat\a}
k_{\hat\a},-\Theta^{\L\hat\a} k_{\hat\a})\ ,
\end{equation}
i.e.\ it selects the generators of the local gauge group $G$. The embedding tensor itself is a spurionic object, which means that it is formally considered to transform as defined by its index structure (adjoint $\times$ fundamental) such that  $Sp(\nv+1)$-covariance is restored in the Lagrangian. In this way, electric and magnetic gaugings are treated on the same footing. Choosing a specific value for the embedding tensor then fixes the gauge group $G$ and breaks the global symmetry $G_0$ to $G\times H$, where $H$ is the maximal commutant of $G$. Consistency of the embedding tensor projection onto the local subset requires that the generators $Y_\Lambda$ form a closed subalgebra $G$ of $G_0$. This is ensured by the quadratic constraint
\begin{equation}\label{embedding_constraint_closure}
f^{\hat\gamma}_{\hat{\alpha}\hat{\beta}} \Theta_{M}^{\hat{\alpha}} \Theta_{N}^{\hat{\beta}} + \Theta_{M}^{\hat{\alpha}} (k_{\hat{\alpha}})_N^{\phantom{N}P} \Theta_{P}^{\hat{\gamma}} = 0 \ ,
\end{equation}
where  $f^{\hat\gamma}_{\hat{\alpha}\hat{\beta}}$ are the structure constants of the global symmetry group $G_0$. In addition, supersymmetry imposes a linear constraint
\begin{equation}\label{embedding_constraint_linear}
 \Theta_{(\Lambda}^{\hat{\a}} (k_{\hat{\a}})_{\Sigma\Xi)} = 0 \ .
\end{equation}
It is also important to note that the requirement of mutually local charges is expressed as an additional constraint on the embedding tensor
\begin{equation}
 \label{embedding_constraint_q}
 \Omega^{\Lambda\Sigma} \Theta_{\Lambda}^{\hat{\alpha}}
 \Theta_{\Sigma}^{\hat{\beta}}\ =\ \Theta^{I[\hat{\alpha}}
 \Theta_{I}^{\phantom{I}\hat{\beta}]}\ =\ 0 \ ,
\end{equation}
where $\Omega^{\Lambda\Sigma}$ is the inverse $Sp(n_{\rm v}+1)$ metric.

In principal, the gauge group $G$ that is selected by the embedding tensor can be Abelian or non-Abelian. In $\cN=2$ supersymmetry any non-Abelian gauge group $G$ always has a Coulomb branch, where the scalars $t^i$ in the adjoint representation have a vacuum expectation value which breaks $G\to [U(1)]^{\rm rank(G)}$. As we will argue in more detail in the next section, partial supersymmetry breaking requires charged hypermultiplets, but not non-Abelian vector multiplets. Thus, for the purposes of this paper we can go far out on the Coulomb branch and safely integrate out all massive vector multiplets, leaving an Abelian theory with charged hypermultiplets at low energies.

For an Abelian theory, no isometries on $\M_{\rm v}$ are gauged and the non-trivial Killing vectors -- denoted by $k_{\lambda}$ -- act only on $\M_{\rm h}$ (see, for instance, \cite{Galicki:1986ja,D'Auria:1990fj}). This immediately implies that the constraints \eqref{embedding_constraint_closure} and \eqref{embedding_constraint_linear} are  trivially satisfied and we only have to impose \eqref{embedding_constraint_q}.
The gauge transformation of the scalar fields $q^u$ in the hypermultiplets then takes the form:
\begin{equation}\begin{aligned}
\delta_\alpha q^u \ &= \ \alpha^\Lambda \Theta_\Lambda^{~~\lambda}
k_{\lambda}^u(q)  \ ,
\end{aligned}\end{equation}
where $k_{\lambda}^u(q)$ are the components of the Killing vectors $k_{\lambda}$, and $\alpha^\Lambda$ are the transformation parameters. In the kinetic terms for the scalars $q^u$ the ordinary derivative is replaced by the covariant derivative
\begin{equation}\begin{aligned}\label{d2}
\partial_\mu q^u\to D_{\mu} q^u &= \pt_{\mu} q^u - A^{~~M}_{\mu} Y_M
q^u\\
&= \pt_{\mu}  q^u - A^{~~\L}_{\mu} \Theta_\L^{~~\lambda} k_{\lambda}^u + B_{\mu\L} \Theta^{\L~{\lambda}} k_{\lambda}^u \ ,
\end{aligned}\end{equation}
while the derivatives of the $t^i$ are unchanged. Inserting the replacement \eqref{d2} into the Lagrangian \eqref{sigmaint} introduces both electric and magnetic vector fields. This upsets the counting of degrees of freedom and leads to unwanted equations of motions. Therefore, the Lagrangian has to be carefully augmented by a set of two-form gauge potentials $B_{\mu\nu}^M$ with couplings that keep supersymmetry and gauge invariance intact. As we do not need these couplings in this paper, we refer the interested reader to the literature for further details~\cite{deWit:2002vt,deWit:2005ub,deVroome:2007zd}.

An analysis of the symplectic extension of the gauged $\cN=2$ supergravity Lagrangian in $D=4$ including electric and magnetic charges has been carried out in \cite{Dall'Agata:2003yr,Sommovigo:2004vj,D'Auria:2004yi}.\footnote{The case of global $\cN=2$ supersymmetry has been studied in the embedding tensor formalism in \cite{deVroome:2007zd}.} We are specifically interested in the scalar part of supersymmetry variations, i.e.\
\begin{eqnarray}\label{susytrans2}
\delta_\epsilon \Psi_{\mu {\cal A}} &=& D_\mu \epsilon^*_{\cal A} - S_{\cal AB} \gamma_\mu \epsilon^{\cal B} + \ldots \, ,\nonumber\\
\delta_\epsilon \lambda^{i {\cal A}} &=& W^{i{\cal AB}}\epsilon_{\cal B}+\ldots \, ,\\
\delta_\epsilon \zeta_{\alpha} &=& N_\alpha^{\cal A} \epsilon_{\cal A}+\ldots \, ,\nonumber
\end{eqnarray}
where the ellipses indicate further terms which vanish in a maximally symmetric ground state. $\gamma_\mu$ are Dirac matrices and $\epsilon^{\cal A}$ is the $SU(2)$ doublet of spinors parameterising the $\cN=2$ supersymmetry transformations.\footnote{Note that the $SU(2)$ R-symmetry acts as the $Sp(1)$ introduced above on the quaternionic-K\"ahler manifold.}
$S_{\cal AB}$ is the mass matrix of the two gravitini, while $W^{i {\cal AB}}$ and $N_\alpha^{\cal A}$ are related to the mass matrices of the spin-$\tfrac12$ fermions, cf.\ \eqref{massmatrices}. The symplectic extensions of these expressions in the embedding tensor formalism are given by are
\begin{eqnarray}\label{susytrans3}
S_{\cal AB} &=& \tfrac{1}{2} \e^{K^{\rm v}/2} {V}^\Lambda \Theta_\Lambda^{~~\lambda} P_{\lambda}^x
(\sigma^x)_{\cal AB} \ ,\nonumber\\
W^{i{\cal AB}}
&=& \mathrm{i} \e^{K^{\rm v}/2} g^{i\bar \jmath}\,
(\nabla_{\bar \jmath}\bar {V}^\Lambda) \Theta_\Lambda^{~~\lambda} P_{\lambda}^x (\sigma^x)^{\cal AB}
\ ,\\
N_\alpha^{\cal A}
&=& 2 \e^{K^{\rm v}/2} \bar {V}^\Lambda \Theta_\Lambda^{~~\lambda} {\cal U}^{\cal A}_{\alpha u} k^u_{\lambda}
\ .\nonumber
\end{eqnarray}
Let us explain the notation used in equations.~\eqref{susytrans3}. The matrices $(\sigma^x)_{\cal AB}$ and $(\sigma^x)^{\cal AB}$ are found by applying the $SU(2)$ metric $\varepsilon_{\cal{AB}}$ (and its inverse) to the standard Pauli matrices $(\sigma^x)_{\cal A}^{~~\cal B}$, $x=1,2,3,$ and are given in Appendix~\ref{section:sg}. ${V}^\Lambda$ is the holomorphic symplectic vector defined by ${V}^\Lambda = (X^I,{\cal F}_I)$ and its K\"ahler covariant derivative is defined as $\nabla_i V^\Lambda = \partial_i V^\Lambda +K^{\rm v}_i V^\Lambda$, with $ K^{\rm v}_i = \partial_i K^{\rm v}$. ${\mathcal U}^{\mathcal A\alpha} = {\mathcal U}_u^{\mathcal A\alpha} \diff q^u$ is the vielbein on the quaternionic-K\"ahler manifold ${\M}_{\rm h}$ and is related to the metric $h_{uv}$ via
\begin{equation}\label{Udef}
h_{uv} \diff q^u \diff q^v = {\mathcal U}^{\mathcal A\alpha} \varepsilon_\mathcal{AB} \mathcal C_{\alpha \beta} \mathcal U^{\mathcal B\beta} \ ,
\end{equation}
where $\mathcal C_{\alpha \beta}$ is the $Sp(\nh)$ invariant metric. ${\mathcal U}^{\mathcal A\alpha}$ satisfies the reality condition
\begin{equation}\label{Ureal}
{\mathcal U}_{\mathcal A\alpha} = \varepsilon_\mathcal{AB} \mathcal C_{\alpha \beta} \mathcal U^{\mathcal B\beta} = ({\mathcal U}^{\mathcal A\alpha})^* \ .
\end{equation}
Finally, $P^x_\lambda, x=1,2,3$ is a triplet of Killing prepotentials defined as
\begin{equation}\label{Pdef}
-2 k^u_\lambda\,K_{uv}^x \ = \  \nabla_v P_\lambda^x\ ,
\end{equation}
where $\nabla_v$ is the $Sp(1)$ covariant derivative, $k_\lambda$ are the isometries on the quaternionic-K\"ahler manifold and $K_{uv}^x$ is the triplet of covariantly constant hyper-K\"ahler two-forms.

Given the supersymmetry variations \eqref{susytrans2} and \eqref{susytrans3}, a Ward identity leads to the general formula for the classical scalar potential $V$ \cite{Cecotti:1984wn,D'Auria:2001kv}:
 \begin{equation} \label{potential_identity}
 V \delta^{\cal A}_{\cal B} = -12 S_{\cal BC} \bar S^{\cal AC} + g_{i\bar \jmath} W^{i \cal {AC}} W^{\bar \jmath}_{\cal BC}
+ 2 N_\alpha^{\cal A} N_{\cal B}^\alpha \ ,
\end{equation}
and it has been argued that this expression holds true in the presence of magnetic charges \cite{Michelson:1996pn,deWit:2005ub}.

\section{Partial Supersymmetry Breaking}\label{section:vectors}

Spontaneous $\cN=2\to \cN=1$ supersymmetry breaking in a Minkowski or AdS ground state requires that for one linear combination
of the two spinors $\epsilon^{\cal A}$ parameterising the supersymmetry transformations, say $\epsilon^{\cal A}_1$, the variations of the fermions given in \eqref{susytrans2} vanish, i.e.\ $\delta_{\epsilon_1} \lambda^{i {\cal A}} = \delta_{\epsilon_1} \zeta_\alpha = \delta_{\epsilon_1} \Psi_{\mu {\cal A}} =0$ (see e.g.\ \cite{Louis:2002vy,Gunara:2003td} for a review). Furthermore, in a supersymmetric Minkowski or AdS background the supersymmetry parameter obeys the Killing spinor equation\footnote{Note that the index of $\epsilon^*_{1\,\cal A}$ is not lowered with $\varepsilon_{\cal AB}$ but $\epsilon^*_{1\,\cal A}$ is related to $\epsilon^A_1$ just by complex conjugation. $|\mu|$ is related to the cosmological constant via $\Lambda = - 3 |\mu|^2$, as we compute in Appendix~\ref{section:stability}, while the phase of $\mu$ is unphysical.}
\begin{equation}
 D_\mu \epsilon^*_{1\,\cal A} = \tfrac12 \mu \gamma_\mu
 \epsilon^*_{1\,\cal A} \ .
\end{equation}
The requirement of a maximally symmetric ground state ensures that the terms which are indicated by the ellipses in \eqref{susytrans2} automatically vanish, so that one is left with\footnote{For a recent discussion of flux compactifications and partially supersymmetric domain walls in $\cN=2$ supergravity, see \cite{Smyth:2009fu}.}
\begin{equation} \label{N=1conditions}
W_{i\cal AB}\, \epsilon^{\cal B}_1\ =\ 0\ =\ N_{\alpha \cal A}\,
\epsilon^{\cal A}_1 \ ,\qquad \textrm{and}\qquad  S_\mathcal{AB}\,
\epsilon^{\cal B}_1\ =\  \tfrac12 \mu \epsilon^*_{1\,\cal A} \ .
\end{equation}
Here we have chosen to write the parameter of the unbroken $\cN=1$ supersymmetry $\epsilon_1$ as a vector $\epsilon^{\cal A}_1$ in the space of $\cN=2$ parameters. For the second, broken generator, which we denote by $\epsilon^{\cal A}_2$, we should have
\begin{equation} \label{N=1conditions2}
W_{i\cal AB}\, \epsilon^{\cal B}_2 \neq 0\qquad \textrm{or}\qquad
N_{\alpha \cal A}\, \epsilon^{\cal A}_2\neq 0\ ,\qquad \textrm{and}\qquad  S_\mathcal{AB}\,
\epsilon^{\cal B}_2\ \ne\  \tfrac12 \mu' \epsilon^*_{2\,\cal A} \ ,
\end{equation}
for any $\mu'$ that obeys $|\mu'|=|\mu|$, i.e.\ only differs from $\mu$ by an unphysical phase.

Before we attempt to solve \eqref{N=1conditions} and \eqref{N=1conditions2} let us assemble a few more facts. A necessary condition for the existence of an $\cN=1$ ground state is that the two eigenvalues $m_{\Psi_1}$ and $m_{\Psi_2}$ of the gravitino mass matrix $S_{\cal AB}$ are non-degenerate, i.e.\ $m_{\Psi_1}\neq m_{\Psi_2}$. In a Minkowski ground state one also needs $m_{\Psi_1}=0$ or, in other words, one of the two gravitini has to become massive, while the second one stays massless, cf.\ \eqref{N=1conditions} and \eqref{N=1conditions2}. Furthermore, the unbroken $\cN=1$ supersymmetry implies that the massive gravitino has to be a member of an entire $\cN=1$ massive spin-$3/2$ multiplet, which has the spin content $s=(3/2,1,1,1/2)$. This means that two vectors, say $A_\mu^0, A_\mu^1$ and a spin-$1/2$ fermion $\chi$ have to become massive, in addition to the gravitino. Therefore, the would-be Goldstone fermion (the Goldstino), which gets eaten by the gravitino, is accompanied by two would-be Goldstone bosons (the sGoldstinos) \cite{Ferrara:1983gn}. The minimum field content of the massive spin-$3/2$ multiplet in terms of massless ${\cal N}=1$ multiplets is then one spin-$3/2$ multiplet, one vector multiplet and one chiral multiplet. Naively, one might think that both the $\cN=1$ vector and chiral multiplet come from $\cN=2$ vector multiplets in a non-Abelian theory, without the need for additional charged hyperscalars. However, vector multiplet scalars are singlets under the $SU(2)$ R-symmetry of $\cN=2$ supergravity and therefore cannot give rise to a mass splitting between the gravitini \cite{Ferrara:1985gj}. In an Abelian theory, on the other hand, the sGoldstinos have to be `recruited' out of a charged $\cN=2$ hypermultiplet, while the need for two gauge bosons implies that at least one $\cN=2$ vector multiplet has to be part of the spectrum. Thus, the minimal $\cN=2$ spectrum which allows for the possibility of a spontaneous breaking to $\cN=1$ consists of the $\cN=2$ supergravity multiplet, one hypermultiplet and one vector multiplet.
In geometric terms, the presence of two sGoldstinos in the hypermultiplet sector means that ${\M}_{\rm h}$ has to admit two commuting isometries, say $k_1^u,k_2^u$, and that these isometries have to be gauged \cite{Fre:1996js}. The definition \eqref{Pdef} then implies that we need to have two non-zero Killing prepotentials $P_1^x, P_2^x$ in the ground state. Furthermore, these prepotentials must not be proportional to each other because otherwise we could take linear combinations of $k_1^u$ and $k_2^u$ such that one combination has vanishing prepotentials. In Section~\ref{section:magnetic_vectors} we verify that two gauged Killing vectors with non-aligned Killing prepotentials are necessary for partial supersymmetry breaking to appear and also discuss the case of more than two gauged Killing vectors. For an $\cN=1$ vacuum in $\cN=2$ supergravity, one can infer stability using the Witten-Nester argument for positive energy \cite{Cecotti:1984wn}. In Appendix~\ref{section:stability} we analyse the scalar potential, showing that a Minkowski or AdS background with $\cN=1$ supersymmetry is automatically stable and obeys the Breitenlohner-Freedman bound \cite{Breitenlohner:1982jf}.

\subsection{The Electric No-go Theorem} \label{section:no-go}
After this initial discussion of the necessary ingredients, let us now discuss the obstructions to spontaneous $\cN=2$ to $\cN=1$ supersymmetry breaking. Using the superconformal tensor calculus, Cecotti et al.\ showed that an $\cN=2$ gauged supergravity with only electric charges cannot have an $\cN=1$ Minkowski ground state \cite{Cecotti:1984rk,Cecotti:1984wn}. More precisely, it was shown that in this case the gravitino mass matrix is proportional to the unit matrix and hence is degenerate. This implies that the ground state either has the full $\cN=2$ supersymmetry or none at all, ruling out the possibility of spontaneous partial supersymmetry breaking.
We shall now review this no-go theorem with the help of the embedding tensor formalism, without using superconformal tensor calculus.
For purely electric gaugings, it turns out that the no-go theorem follows from the gravitino and gaugino variations alone. The hyperino
equation gives additional constraints on the hypermultiplet sector, and we postpone its discussion to Section~\ref{section:hypermultiplets}.

Assume that we are at a point $X^I_0$ in the vector multiplet moduli space and at a point $q^u_0$ in the quaternionic-K\"ahler manifold at which supersymmetry is broken to $\cN=1$ and the conditions \eqref{N=1conditions} hold. For simplicity, we shall drop the subscript and simply denote this point by $X^I$ and $q^u$. The gravitino equation in \eqref{N=1conditions} for electric gaugings is given by
\begin{equation}\label{S_N=1}
  S_\mathcal{AB}\, \epsilon_1^{\cal B} = \tfrac{1}{2} \e^{K^{\rm v}/2} X^I
  \Theta_I^{\phantom{I}\lambda} P_\lambda^x \sigma^x_\mathcal{AB}
  \epsilon_1^{\cal B} = \tfrac12 \mu \epsilon^*_{1\,\cal A} \ .
\end{equation}
The (complex conjugate) of the gaugino variation in \eqref{N=1conditions} leads to
\begin{equation}\begin{aligned}\label{W_N=1}
 W_{i\cal AB}\, \epsilon_1^{\cal B} &= \iu \e^{K^{\rm
 v}/2}(\nabla_i X^I)
 \Theta_I^{\phantom{I}\lambda} P_\lambda^x \sigma^x_\mathcal{AB}
 \epsilon_1^{\cal B} \\
&=\iu
 \e^{K^{\rm v}/2}  (\partial_i X^I) \Theta_I^{\phantom{\Lambda}\lambda}
 P_\lambda^x \sigma^x_\mathcal{AB} \epsilon_1^{\cal B}
+  \iu K^{\rm v}_i \mu \epsilon^*_{1\,\cal A}  = 0 \ ,
\end{aligned}\end{equation}
where in the second line we have used $\nabla_i X^I = \partial_i X^I + K^{\rm v}_i X^I$ and inserted the gravitino equation \eqref{S_N=1}. Note that in total \eqref{S_N=1} and \eqref{W_N=1} give $2(\nv+1)$ equations to solve. Let us now specialise to a frame where a prepotential exists. We can then express $X^I$ in terms of special coordinates as $X^I = (1, t^i)$ and we find that the gaugino equation \eqref{W_N=1} simplifies to
\begin{equation} \label{nogo_Theta_i}
 \Theta_i^{\phantom{i}\lambda} P_\lambda^x \sigma^x_\mathcal{AB}
 \epsilon_1^{\cal B} = - \e^{-K^{\rm v}/2} \mu K^{\rm v}_i \epsilon^*_{1\,\cal A}  \ .
\end{equation}
Inserting this back into the gravitino equation \eqref{S_N=1} yields
\begin{equation}\label{noint}
 \Theta_0^{\phantom{0}\lambda} P_\lambda^x \sigma^x_\mathcal{AB}
 \epsilon_1^{\cal B}\ =\ \e^{-K^{\rm v}/2} \mu\, (1+t^i K^{\rm v}_i)\, \epsilon^*_{1\,\cal A}  \ .
\end{equation}
From the definition of the K\"ahler potential \eqref{Kdef} one derives the identity $X^I K^{\rm v}_I = -1$, which in special coordinates $X^I = (1, t^i)$ reads $1+t^i K^{\rm v}_i= - K^{\rm v}_0 $. This further simplifies \eqref{noint} to give
\begin{equation}\label{noint2}
 \Theta_0^{\phantom{0}\lambda} P_\lambda^x \sigma^x_\mathcal{AB}
 \epsilon_1^{\cal B} = - \e^{-K^{\rm v}/2} \mu K^{\rm v}_0 \epsilon^*_{1\,\cal A}  \ ,
\end{equation}
which allows us to combine \eqref{nogo_Theta_i} and \eqref{noint2} into the $2(\nv+1)$ equations
\begin{equation}\label{noend}
 \Theta_I^{\phantom{I}\lambda} P_\lambda^x \sigma^x_\mathcal{AB}\,
 \epsilon_1^{\cal B} = - \e^{-K^{\rm v}/2} \mu K^{\rm v}_I  \epsilon^*_{1\,\cal A}  \ .
\end{equation}
To summarise, by using the existence of the special coordinates  $X^I = (1, t^i)$ we have been able to rewrite the original $2(\nv+1)$ equations arising from the gravitino \eqref{S_N=1} and gaugino \eqref{W_N=1} variations in a compact manner \eqref{noend}.

If we now consider a Minkowski vacuum, setting $\mu=0$, the expression $\sigma^x_\mathcal{AB} \epsilon_1^{\cal B}$ is the only complex quantity appearing in \eqref{noend}. Therefore, we can use \eqref{noend} with $\mu=0$ and its complex conjugate to find
\begin{equation}\label{Theta1}
  \Theta_I^{\phantom{I}\lambda} P_\lambda^x = 0 \ .
\end{equation}
If we then insert \eqref{Theta1} back into the matrices appearing in the supersymmetry transformations \eqref{susytrans3}, we see that $S_{\cal AB}$ (and $W_{i{\cal AB}}$) identically vanish, and thus partial supersymmetry breaking is not possible, i.e.\ we have recovered the original no-go theorem \cite{Cecotti:1984rk}. The important step in this derivation was using the existence of a prepotential and the special coordinates $X^I = (1, t^i)$ to find $\nv$ independent equations in \eqref{nogo_Theta_i}. Therefore, this no-go theorem might be circumvented by using a symplectic frame in which no prepotential exists at the $\cN=1$ point. It turns out that this is possible, and the first examples of spontaneous partial supersymmetry breaking used precisely such frames where the prepotential does not exist \cite{Ferrara:1995gu,Ferrara:1995xi,Fre:1996js}. On the other hand, such symplectic frames are related to the standard one by a symplectic transformation which just rotates electric and magnetic charges into each other. Therefore, in the following  we still assume the existence of a prepotential but generalise our discussion by allowing for both electric and magnetic charges. This covers all possible gauged supergravities and in particular the examples mentioned above. In the next section, we show that this generalisation indeed gives rise to the possibility of spontaneous partial supersymmetry breaking.

\subsection{A Way Out - Magnetic Fluxes} \label{section:magnetic_vectors}
We shall now repeat the discussion of Section~\ref{section:no-go} with magnetic gaugings included. We will also discuss partial supersymmetry breaking to both Minkowski and AdS vacua, i.e.\ we keep $\mu$ nonzero in \eqref{N=1conditions}. First, we note that the condition which comes from the vanishing of the gaugino variation \eqref{W_N=1}, now with electric and magnetic gaugings, gives rise to
\begin{equation}\label{N=1_condition_vectorsI}
\e^{K^{\rm v}/2}(\partial_i X^I \Theta_I^{\phantom{I}\lambda} - \partial_i {\cal F}_I \Theta^{I\lambda}) P_\lambda^x \sigma^x_\mathcal{AB} \epsilon_1^{\cal B} +  K^{\rm v}_i \mu \epsilon^*_{1\,\cal A}  = 0 \ ,
\end{equation}
where the second term in the brackets is due to the presence of magnetic charges $\Theta^{I\lambda}$. Contracting \eqref{N=1_condition_vectorsI} with $t^i$ and adding it to  $2 S_\mathcal{AB} \epsilon_1^{\cal B}= \mu \epsilon_{1\,\cal A}^*$ we arrive at
\begin{equation}\begin{aligned}\label{int}
\e^{-K^{\rm v}/2} \mu (1+t^i K^{\rm v}_i) \epsilon_{1\,\cal A}^* &= (X^I \Theta_I^{\phantom{I}\lambda} - {\cal F}_I \Theta^{I\lambda}) P_\lambda^x \sigma^x_\mathcal{AB} \epsilon_1^{\cal B} - t^i (\Theta_i^{\phantom{i}\lambda} - {\cal F}_{iJ} \Theta^{J\lambda}) P_\lambda^x \sigma^x_\mathcal{AB} \epsilon_1^{\cal B}\\
&= (\Theta_0^{\phantom{0}\lambda} - {\cal F}_{0J} \Theta^{J\lambda}) P_\lambda^x \sigma^x_\mathcal{AB} \epsilon_1^{\cal B} \ .
\end{aligned}\end{equation}
Using again $1+t^i K^{\rm v}_i= - K^{\rm v}_0 $ in \eqref{int} and combining it with \eqref{N=1_condition_vectorsI} yields $2(\nv+1)$ equations, replacing the conditions \eqref{noend} of the previous section:
\begin{equation} \label{condition_vectors}
 (\Theta_I^{\phantom{I}\lambda} - {\cal F}_{IJ} \Theta^{J\lambda}) P_\lambda^x \sigma^x_\mathcal{AB} \epsilon_1^{\cal B} = - \e^{-K^{\rm v}/2} \mu K^{\rm v}_I \epsilon_{1\,\cal A}^*  \ .
\end{equation}
These equations give conditions on the embedding tensor and on the prepotential. However, in order to ensure that the second supersymmetry is broken the conditions \eqref{condition_vectors} should not simultaneously hold for the second supersymmetry generator
\begin{equation} \label{susy_generator_2}
 \epsilon_2^{\cal A} = (\varepsilon_{\cal AB} \epsilon_1^{\cal B})^* \ .
\end{equation}
Inserting \eqref{susy_generator_2} into \eqref{condition_vectors}, we arrive at the additional condition
\begin{equation} \label{condition_vectors_2}
 (\Theta_I^{\phantom{I}\lambda} - \bar{\cal F}_{IJ} \Theta^{J\lambda})
 P_\lambda^x \sigma^x_\mathcal{AB} \epsilon_1^{\cal B} \ne \e^{-K^{\rm
 v}/2} \bar \mu' \bar K^{\rm v}_I \epsilon_{1\,\cal A}^*  \quad
 \textrm{for some} \ I \ ,
\end{equation}
for any $\mu'$ that obeys $|\mu'|=|\mu|$.

\subsubsection{Minkowski Vacua}

Let us proceed by first analysing Minkowski vacua ($\mu=0$). For this case \eqref{condition_vectors} and \eqref{condition_vectors_2} simplify to
\begin{subequations}\label{condition_vectors_M}
 \begin{eqnarray}
 (\Theta_I^{\phantom{I}\lambda} - {\cal F}_{IJ} \Theta^{J\lambda})\,
 P_\lambda^x \sigma^x_\mathcal{AB} \epsilon_1^{\cal B}\ &=\ 0  \quad
 \textrm{for all} \ I \ ,\quad \label{condition_vectors_M1} \\
 (\Theta_I^{\phantom{I}\lambda} -  \bar{\cal F}_{IJ} \Theta^{J\lambda})\, P_\lambda^x
 \sigma^x_\mathcal{AB} \epsilon_1^{\cal B}\ &\ne\ 0  \quad \textrm{for some} \ I \ . \label{condition_vectors_M2}
\end{eqnarray}
\end{subequations}
The crucial point is that the existence of an ${\cal N}=1$ vacuum requires that there is a set of charges for which \eqref{condition_vectors_M1} vanishes while \eqref{condition_vectors_M2} does not.
If \eqref{condition_vectors_M2} were also to vanish for all $I$, then the vacuum would preserve the full ${\cal N}=2$ supersymmetry.\footnote{For the subset of $I$ for which \eqref{condition_vectors_M2} does also vanish, we can add and subtract the equations \eqref{condition_vectors_M1} and \eqref{condition_vectors_M2} such that $\sigma^x_\mathcal{AB} \epsilon_1^{\cal B}$ is the only complex quantity in the resulting equations. Analogously to the discussion above \eqref{Theta1}, this then leads to $(\Theta_I^{\phantom{I}\lambda} - {\cal F}_{IJ} \Theta^{J\lambda}) P^x_\lambda = 0$. If this is the case for all $I$, we have $S_{\cal AB}=0$ and thus an $\cN=2$ vacuum.}
On the other hand, for an $\cN=1$ vacuum it is sufficient to find that for some $I$ \eqref{condition_vectors_M2} does not vanish.
Let us also reiterate that \eqref{condition_vectors_M1} and \eqref{condition_vectors_M2} do not have to hold over all of field space but only at the $\cN=1$ point. As ${\cal N}=1$ supersymmetry is preserved, one can show that this point is a minimum of the potential, see Appendix~\ref{section:stability}.

Before solving \eqref{condition_vectors_M} let us first recall that we must have two commuting isometries $k_1$ and $k_2$ on ${\M}_{\rm h}$, as discussed at the beginning of Section~\ref{section:vectors}, and that at the $\cN=1$ point the corresponding Killing prepotentials $P^x_1$ and $P^x_2$ are both non-vanishing and not proportional to each other.
Consider \eqref{condition_vectors_M} with just one gauged isometry, say $k_1$. In this case \eqref{condition_vectors_M1} factorises into two parts i.e. either $(\Theta_I^{\phantom{I}1} - {\cal F}_{IJ} \Theta^{J1})$ or $P_1^x \sigma^x_\mathcal{AB} \epsilon_1^{\cal B}$ must vanish. However, from \eqref{condition_vectors_M2} we see that both of these expressions have to be non-zero. Therefore, for one gauged isometry we can only have ${\cal N}=2$ or ${\cal N}=0$. We shall first study the case with two gauged isometries and discuss the case with more gauged isometries later.
As our analysis is local, we can choose a convenient $SU(2)$ frame to further simplify \eqref{condition_vectors_M}. In particular, we can choose $P^x_1$ and $P^x_2$ to lie in the $x=1,2$ plane. Furthermore, we can make use of the complex combination
\begin{equation}\label{Pc}
 P_{1,2}^\pm = P_{1,2}^1 \pm \iu P_{1,2}^2 \ .
\end{equation}

We will now construct an embedding tensor $\Theta_\Lambda^{1,2}$ such that in this $SU(2)$ frame the supersymmetry generated
by $\epsilon_1^{\cal A} = (\epsilon_1^1,0)$ is unbroken. Using \eqref{Pc}, \eqref{condition_vectors_M} becomes
\begin{subequations}
\begin{eqnarray}
 P^-_1(\Theta_I^{\phantom{I}1} - {\cal F}_{IJ} \Theta^{J1}) + P^-_2
 (\Theta_I^{\phantom{I}2} - {\cal F}_{IJ} \Theta^{J2})
&=& 0 \quad \textrm{for all} \ I \ , \label{condition_vectors_Msolve1} \\
 P^-_1(\Theta_I^{\phantom{I}1} - \bar{\cal F}_{IJ}
 \Theta^{J1})+P^-_2 (\Theta_I^{\phantom{I}2} - \bar{\cal F}_{IJ}
 \Theta^{J2})
& \ne & 0 \quad \textrm{for some} \ I  \ . \label{condition_vectors_Msolve2}
\end{eqnarray}
\end{subequations}
Applying the elementary identity
\begin{equation} \label{identity_v}
 \Im({\cal F}_{IJ}\Phi^J) - {\cal F}_{IJ} \Im \Phi^J = (\Im{\cal F})_{IJ} \bar \Phi^J \ ,
\end{equation}
which holds for any complex vector $\Phi^I$,
we can solve \eqref{condition_vectors_Msolve1} in
terms of an arbitrary
complex vector $\C^I$ by choosing
\begin{equation}\label{solution_embedding_tensor}
\begin{aligned}
\Theta_I^{\phantom{I}1} = & -\Im(P^+_2 {\cal F}_{IJ}\C^J ) \ , \qquad  \Theta^{I1} = & -\Im (P^+_2\C^I) \ , \\
\Theta_I^{\phantom{I}2} = & \quad \Im(P^+_1 {\cal F}_{IJ}\C^J ) \ , \qquad  \Theta^{I2} = & \quad \Im (P^+_1 \C^I) \ ,
\end{aligned}
\end{equation}
where the Killing prepotentials and ${\cal F}_{IJ}$ are evaluated at the local $\cN=1$ minimum.
Note that since $P^x_1$ and $P^x_2$ are not aligned, the expression \eqref{condition_vectors_Msolve2} cannot vanish for any non-zero $\C^I$.

We also need to enforce the mutual locality constraint \eqref{embedding_constraint_q}, which for the case at hand reads
\begin{equation}\label{locality_two_isometries}
 \Theta^{I1}\Theta_I^{\phantom{I}2} - \Theta^{I2} \Theta_I^{\phantom{I}1} = 0 \ .
\end{equation}
If we now insert the solutions \eqref{solution_embedding_tensor} into this constraint, we find a condition on the coefficients  $\C^{I}$:
\begin{equation} \label{constraint_embedding_tensor}
  \bar{\C}^{I} (\Im {\cal F})_{IJ} \C^{J}  = 0 \ .
\end{equation}
In deriving this we have used that $\Im(P^-_1 P^+_2)\ne0$, which holds because the prepotentials $P^x_1$ and $P^x_2$ are not aligned. Therefore, we have found that $\C^{I}$ has to be null with respect to $(\Im {\cal F})_{IJ}$. Since $(\Im {\cal F})_{IJ}$ is of signature $(\nv,1)$, as we later show in \eqref{signature_G}, this constraint can be easily satisfied.
Therefore, we have found that breaking $\cN=2$ to $\cN=1$ supersymmetry is possible.

We can perform a symplectic rotation $\cal S$ given by\footnote{For further details on symplectic transformations in $\cN=2$ supergravity, see Appendix~\ref{section:sg}.}
\begin{equation}
  {\cal S}^\Lambda_{~\Sigma} \ = \left(
    \begin{array}{cc}
      U^I_{~J} & Z^{IJ} \\
      [1mm] W_{IJ} & V^{~J}_I
    \end{array}
  \right) \ ,
\end{equation}
to transform $\Theta^{\Lambda1,2} = (\Theta^{I1,2}, \Theta^{\ 1,2}_I)$ such that we only have electric charges, in other words $\Theta^{I\,1,2}$ vanishes in the rotated frame. We see from \eqref{solution_embedding_tensor} that then also the symplectic vector
\begin{equation}
 \left( \begin{array}{cc}
  \C^I \\ {\cal F}_{IJ} C^J
 \end{array} \right)
 = \tfrac{P^-_1}{\Im(P^+_1 P^-_2)}  
 \left( \begin{array}{cc} 
  \Theta^{I1} \\ \Theta_I^{\ 1}
 \end{array} \right) + \tfrac{P^-_2}{\Im(P^+_1 P^-_2)}  \left( \begin{array}{cc} 
  \Theta^{I2} \\ \Theta_I^{\ 2}
 \end{array} \right) \ , 
\end{equation}
has to become purely electric under ${\cal S}$, i.e.\
\begin{equation}
(U^I_{~J} + Z^{IK} {\cal F}_{KJ}) \C^J = 0 \ , 
\end{equation}
and thus the matrix $U^I_{~J} + Z^{IK} {\cal F}_{KJ}$ is not invertible.
As discussed in \cite{Ceresole:1995jg}, this precisely means that we transform into a symplectic frame where no prepotential exists at the $\cN=1$ point, as demanded by the no-go theorem we reviewed in Section~\ref{section:no-go}.

Let us now consider the case with $n$ gauged commuting isometries. We can always go to a new basis of Killing vectors $k_\lambda$ where there are only three Killing vectors that have $P_{\lambda}^x\ne 0$ at the $\cN=1$ point. Imposing \eqref{condition_vectors_M1} then tells us that at least one combination of the $P^x$ has to vanish and, therefore, there are effectively only two Killing vectors with non-vanishing $P_{\lambda}^x$ at the $\cN=1$ point. We can identify these two Killing vectors with those used above to construct the $\cN=1$ solution. The other Killing vectors do not play a role in the supersymmetry breaking, but could give rise to additional masses as the derivatives of their $P_{\lambda}^x$'s could be non-zero.


The above result is quite surprising. By appropriately choosing the embedding tensor, the conditions for partial $\cN=1$ supersymmetry breaking arising from the gravitino and the gaugino variations can be fulfilled for \emph{any} point on \emph{any} special K\"ahler
manifold ${\M}_{\rm v}$ and for \emph{any} quaternionic-K\"ahler manifold ${\M}_{\rm h}$ that admits two commuting isometries with Killing prepotential $P^x_1$ and $P^x_2$ that are not proportional to each other at the $\cN=1$ point. Of course, we still have to satisfy the non-trivial condition $N_{\alpha \cal A} \epsilon_1^{\cal A} = 0$ of \eqref{N=1conditions}. We shall turn to this issue in Section~\ref{section:hypermultiplets}, where we show that it can be solved for any special quaternionic-K\"ahler manifold.

Before we consider the analysis of AdS vacua, let us discuss a simple example given by the four-dimensional quaternionic-K\"ahler manifold ${\M}_{\rm h}=SO(1,4)/SO(4)$ with arbitrary ${\M}_{\rm v}$.  ${\M}_{\rm h}$ is parameterised by the quaternionic coordinates $(q^0, q^1,q^2, q^3)$ and admits the commuting Killing vectors $k_\lambda=\tfrac{\partial}{\partial q^\lambda}$ for $\lambda=1,2,3$. The Killing prepotentials are given by \cite{Ferrara:1995gu,Ferrara:1995xi}
\begin{equation}\label{prepotentials_ex}
P_\lambda^x = \tfrac{1}{q^0} \delta_\lambda^x \ ,
\end{equation}
which, when inserted into our solution for the embedding tensor components \eqref{solution_embedding_tensor}, yield
\begin{equation}\label{solution_embedding_tensor_ex}
\begin{aligned}
\Theta_I^{\phantom{I}1} = & -\Re( {\cal F}_{IJ}\C^J ) \ , \qquad \Theta^{I1} = & -\Re \C^I \ , \\
\Theta_I^{\phantom{I}2} = &  \quad \Im( {\cal F}_{IJ} \C^J ) \ , \qquad \Theta^{I2} = & \quad \Im \C^I \ .
\end{aligned}
\end{equation}
In this case, it can easily be shown that the hyperino variation $N_{\alpha \cal A}  \epsilon_1^{\cal A} = 0$ is automatically satisfied and we recover the $\cN=1$ vacuum given in \cite{Ferrara:1995gu}. However, the example in \cite{Ferrara:1995gu} was for a specific choice of ${\M}_{\rm v}$, whereas we have just shown that partial supersymmetry breaking is possible for arbitrary ${\M}_{\rm v}$.

\subsubsection{AdS Vacua}

Let us now consider partial supersymmetry breaking in an AdS vacuum, i.e.\ for
$\mu \ne 0$. We again require that there are two commuting Killing vectors with non-aligned
Killing prepotentials and choose an $SU(2)$ frame where $P^x_1$ and $P^x_2$ are in the $x=1,2$ plane. We shall also make use of the identity
\begin{equation}
K^{\rm v}_I = 2\e^{K^{\rm v}} (\Im {\cal F})_{IJ} \bar X^J \ ,
\end{equation}
which follows from the definition of the K\"ahler potential \eqref{Kdef}. We then find that the gaugino conditions \eqref{condition_vectors} simplify and, as a consequence, the first condition for partial supersymmetry breaking is\footnote{The second condition similarly follows from \eqref{condition_vectors_2}.}
\begin{equation}\label{condition_vectors_AdS}
 P^-_1(\Theta_I^{\phantom{I}1} - {\cal F}_{IJ} \Theta^{J1}) + P^-_2
 (\Theta_I^{\phantom{I}2} - {\cal F}_{IJ} \Theta^{J2}) = - 2 \e^{K^{\rm v}/2}\mu (\Im {\cal F})_{IJ} \bar X^J  \ .
\end{equation}
This is just the Minkowski condition \eqref{condition_vectors_Msolve1} with an additional inhomogeneity proportional to $\mu$. If we now again make use of the identity \eqref{identity_v}, the solution to \eqref{condition_vectors_AdS} can be obtained analogously to the Minkowski case \eqref{solution_embedding_tensor}
\begin{equation}\label{solution_embedding_tensor_AdS}
\begin{aligned}
\Theta_I^{\phantom{I}1} = & -\Im({\cal F}_{IJ}(P^+_2 \C_{\rm AdS}^J + \e^{K^{\rm v}/2} \tfrac{\bar \mu}{P^+_1} X^J) ) \ , \\
\Theta^{I1} = & -\Im (P^+_2\C_{\rm AdS}^I + \e^{K^{\rm v}/2} \tfrac{\bar \mu}{P^+_1} X^I)) \ , \\
\Theta_I^{\phantom{I}2} = & \quad \Im({\cal F}_{IJ}(P^+_1 \C_{\rm AdS}^J - \e^{K^{\rm v}/2} \tfrac{\bar \mu}{P^+_2} X^J) ) \ , \\
\Theta^{I2} = & \quad \Im (P^+_1 \C_{\rm AdS}^I- \e^{K^{\rm v}/2} \tfrac{\bar \mu}{P^+_2} X^I) \ ,
\end{aligned}
\end{equation}
where again $\C_{\rm AdS}^I$ is an arbitrary vector. The mutual locality constraint \eqref{constraint_embedding_tensor} now reads 
\begin{equation}\label{constraint_embedding_tensor_AdS_g}
\begin{aligned}
 \bar{\C}_{\rm AdS}^{I} (\Im {\cal F})_{IJ} \C_{\rm AdS}^{J} + \tfrac{|\mu|^2}{2 |P_1|^2 |P_2|^2} = - 2 \tfrac{\Re(P^-_1 P^+_2 )}{\Im(P^-_1 P^+_2 )} \e^{K^{\rm v}/2} \Im \left(\tfrac{\bar \mu}{P^+_1 P^+_2} \bar{\C}_{\rm AdS}^{I} (\Im {\cal F})_{IJ} X^J\right)  \ .
\end{aligned}
\end{equation}
For instance, if we choose the phase of $\C_{\rm AdS}^{I}$ appropriately, the right-hand side of this constraint vanishes and we end up with
\begin{equation}\label{constraint_embedding_tensor_AdS}
 \bar{\C}_{\rm AdS}^{I} (\Im {\cal F})_{IJ} \C_{\rm AdS}^{J} = - \tfrac{|\mu|^2}{2 |P_1|^2 |P_2|^2} \ ,
\end{equation}
which tells us that $\bar{\C}_{\rm AdS}^{I}$ is timelike with respect to $(\Im {\cal F})_{IJ}$. Once again, as $(\Im {\cal F})_{IJ}$ is of signature $(\nv,1)$, cf.\ discussion in \eqref{signature_G}, this condition is easily satisfied. It is straightforward to check that the second condition \eqref{condition_vectors_2} is automatically satisfied and we find that the breaking from $\cN=2$ to $\cN=1$ is possible for any solution in \eqref{solution_embedding_tensor_AdS} with non-zero $\C_{\rm AdS}^I$ obeying \eqref{constraint_embedding_tensor_AdS}.
Similarly to the Minkowski case, the discussion for $n$ gauged commuting isometries always reduces to the above, i.e.\ to just two gauged isometries with non-vanishing prepotentials, while the other gauged isometries can only induce mass terms at the $\cN=1$ point.

This concludes our analysis of the gravitino and gaugino variations. We found that in both Minkowski and AdS spacetimes partial supersymmetry breaking does not constrain the special K\"ahler geometry, but essentially only imposes a condition on  the structure of the embedding tensor. In other words, this is a constraint on the choice of gauge vectors. In addition, two commuting isometries have to exist on the scalar field space $\M_{\rm  h}$. This imposes additional constraints in the hypermultiplet sector, to which we now turn.


\section{The Hypermultiplet Sector}
\label{section:hypermultiplets}

In this section we shall analyse the additional constraints on ${\cal N}=1$ vacua which arise in the hypermultiplet sector. As we stated above, we need to have two commuting isometries on $\M_{\rm  h}$. This is certainly not satisfied on a generic quaternionic-K\"ahler
manifold and so $\M_{\rm  h}$ is constrained from the outset by this requirement. We then need that the unbroken $N=1$ supersymmetry is also respected by the hyperino variation, which is equivalent to solving $N^\alpha_{\cal A} \epsilon^{\cal A}_1=0$. It is difficult to analyse this condition on an arbitrary $\M_{\rm  h}$ which admits two isometries. To proceed, we shall focus our attention on a specific subclass of quaternionic-K\"ahler manifolds - the so called `special quaternionic-K\"ahler manifolds' - which are known to arise at the string tree-level of Calabi-Yau compactifications of type II string theories \cite{Cecotti:1988qn,Ferrara:1989ik}. Beyond their interest in string compactifications, we have chosen to concentrate on this specific subclass as they have a large number of isometries. Before we look for a solution of
$N^\alpha_{\cal A} \epsilon^{\cal A}_1=0$, let us briefly recall some features of special quaternionic-K\"ahler manifolds that will prove useful in the following.

\subsection{Special Quaternionic Spaces} \label{section:special_quat}
Special quaternionic-K\"ahler manifolds are quaternionic-K\"ahler manifolds which contain a $(2\nh-2)$--dimensional submanifold $\M_{\rm  sk}$ that is special K\"ahler. As mentioned previously, they arise in Calabi-Yau compactifications of type II string theories and their construction is known as the c-map \cite{Cecotti:1988qn,Ferrara:1989ik}. In type IIA $\M_{\rm  sk}$ is spanned by the complex-structure deformations of the Calabi-Yau, while in type IIB it is spanned by the K\"ahler deformations. In the following we will not distinguish between the two cases and always denote the coordinates of $\M_{\rm  sk}$ by the complex $z^a, a=1,\ldots,\nh-1$. The other hypermultiplet scalars are the dilaton $\phi$, the axion $\ax$ and  $2\nh$ scalars arising in the
Ramond-Ramond sector which we denote by the real $\xi^A, \tilde\xi_A, A=1,\ldots,\nh$. Due to their Ramond-Ramond origin the couplings of the $\xi^A, \tilde\xi_A$ are very restricted. Furthermore, the dilaton $\phi$ and the axion $\ax$ have universal properties that are independent of the chosen compactification manifold. Together these scalars define a $G$-bundle over $\M_{\rm  sk}$, where $G$ is the semidirect product of a $(2\nh+1)$-dimensional Heisenberg group with $\mathbb{R}$. As a consequence $(2\nh+2)$ independent isometries exist, as we shall discuss further shortly.

The Lagrangian is completely determined in terms of the holomorphic prepotential ${\cal G}$ of the special K\"ahler submanifold. More specifically, the  K\"ahler potential $K^{\rm h}$ of $\M_{\rm  sk}$ is given by
\begin{equation}\label{Khdef}
 K^{\rm h} = - \ln  \iu \left(\bar{Z}^A \mathcal{G}_A - {Z}^A
 \bar{\mathcal{G}}_A \right) \ ,
\end{equation}
where $\mathcal G_A$ denotes the first derivative of the holomorphic prepotential $\mathcal G$ and
$Z^A$ are the homogeneous coordinates, which can be chosen to be $Z^A = (1, z^a)$ in special coordinates. The equivalent of the
gauge field kinetic matrix ${\cal N}_{IJ}$ \eqref{Ndef} is given by
\begin{equation} \label{period_matrix}
\mathcal M_{AB} = \bar{\mathcal G}_{AB} + 2 \iu \frac{(\Im  \mathcal G_{AC}) Z^C (\Im  \mathcal G_{BD}) Z^D}{Z^E (\Im  \mathcal G_{EF}) Z^F} \ ,
\end{equation}
and satisfies
\begin{equation}
\mathcal G_A = \mathcal M_{AB} Z^B \ , \quad \nabla_c \mathcal G_A = \bar{\mathcal M}_{AB} \nabla_c Z^B \ .
\end{equation}

In \cite{Ferrara:1989ik} it was observed that there is a specific parametrisation of the quaternionic vielbein $\mathcal U^{\mathcal A\alpha}$ \eqref{Udef} which turns out to be useful on special quaternionic-K\"ahler manifolds. Specifically, one defines the quaternionic vielbein as\footnote{Our notation follows Ref.~\cite{Cassani:2007pq}.}
\begin{equation} \label{quat_vielbein}
\mathcal U^{\mathcal A\alpha}= \tfrac{1}{\sqrt{2}}
\left(\begin{aligned}
 \bar{u} && \bar{e} && -v && -E \\
\bar{v} && \bar{E} && u && e
\end{aligned}\right) \ ,
 \end{equation}
where the one-forms are defined as
\begin{equation} \label{one-forms_quat}
 \begin{aligned}
  u\ = &\ \iu \e^{K^{\rm h}/2+\phi}Z^A(\diff \tilde\xi_A - \mathcal M_{AB} \diff \xi^B) \ , \\
  v\ = &\ \tfrac{1}{2} \e^{2\phi}\big[ \diff \e^{-2\phi}-\iu (\diff \ax +\tilde\xi_A \diff \xi^A-\xi^A \diff \tilde \xi_A  ) \big] \ , \\
  E^{\,\underline{b}}\ = &\ -\tfrac{\iu}{2} \e^{\phi-K^{\rm h}/2} {\Proj}_A^{\phantom{A}\underline{b}} (\Im  \mathcal G)^{-1\,AB}(\diff \tilde\xi_B - \mathcal M_{BC} \diff \xi^C) \ , \\
  e^{\,\underline{b}} \ = &\ {\Proj}_A^{\phantom{A}\underline{b}} \diff Z^A \ . \\
 \end{aligned}
\end{equation}
In these expressions ${\Proj}_A^{\phantom{A}\underline{b}}$ is defined by
\begin{equation}\label{projector_vielbein}
{\Proj}_A^{\phantom{A}\underline{b}} =({\Proj}_0^{\phantom{0}\underline{b}}, {\Proj}_a^{\phantom{a}\underline{b}}) =(-e_a^{\phantom{a}\underline{b}}Z^a, e_a^{\phantom{a}\underline{b}}) \ ,
\end{equation}
where $e_a^{\phantom{a}\underline{b}}$ is the vielbein of $\M_{\rm sk}$,
i.e.\ it satisfies
$g_{a\bar b} = e_a^{\phantom{a}\underline{b}}
\bar e_{\bar
  b}^{\phantom{a}\bar{\underline{c}}}\delta_{\underline{b}\bar{\underline{c}}},~(
\underline{a},\underline{b} = 1,\ldots,\nv-1$ with $g_{a\bar b})$ being
the metric on $\M_{\rm sk}$.
Note that ${\Proj}_A^{\phantom{A}\underline{b}}$ satisfies ${\Proj}_A^{\phantom{A}\underline{b}} Z^A =0$. It is important to mention that the parametrisation  of the vielbein specified by \eqref{quat_vielbein} and \eqref{one-forms_quat} singles out a particular $SU(2)$ frame on ${\M}_{\rm h}$. As a consequence, any solution of partial supersymmetry breaking that we are going to find will not be $SU(2)$ covariant.

Due to its specific construction, ${\M}_{\rm h}$ has $(2\nh+2)$ isometries which are generated by the following set of Killing vectors
\begin{equation}\label{Killing}
\begin{aligned}
\Kdil  \ &=\ \tfrac{1}{2} \frac{\partial}{\partial \phi} -  \ax \frac{\partial}{\partial \ax} - \tfrac{1}{2} \xi^A \frac{\partial}{\partial \xi^A} - \tfrac{1}{2} \tilde \xi_A \frac{\partial}{\partial \tilde \xi_A} \ , \\
\Kax  \ &=\ - 2 \frac{\partial}{\partial \ax} \ , \\
 \Kxi_A \ &= \ \frac{\partial}{\partial \xi^A} + \tilde \xi_A \frac{\partial}{\partial \ax} \ , \\
\Ktxi^A \ &=\ \frac{\partial}{\partial \tilde \xi_A} - \xi^A \frac{\partial}{\partial \ax} \ .
\end{aligned}
\end{equation}
They act transitively on the $G$-fibre coordinates $(\phi, \ax, \xi^A,\tilde \xi_A)$ and the subset $\{\Kxi_A, \Ktxi^A,\Kax\} $ spans a Heisenberg algebra which is graded with respect to $k_\phi $. The corresponding commutation relations are given by
\begin{equation}\label{isometries_fibre}
\begin{aligned}
 && [\Kdil,\Kax]\  =& \Kax \ , \qquad \qquad
 && [\Kdil,\Kxi_A]\ \ = & \tfrac{1}{2} \Kxi_A \ ,  \\
 && [\Kdil,\Ktxi^A]\  =& \tfrac{1}{2} \Ktxi^A \ , \qquad \qquad
 && [\Kxi_A,\Ktxi^B]\  = & - \delta_A^B \Kax \ ,
\end{aligned}
\end{equation}
while all other commutators vanish.

We shall also need the explicit form of the Killing prepotentials $P^x_\lambda$, which were defined in \eqref{Pdef}. For special quaternionic-K\"ahler manifolds it has been shown that Killing prepotentials take a simple form in terms of the $SU(2)$ connection $\omega^x, x=1,2,3$ \cite{Michelson:1996pn}:
\begin{equation} \label{prepotential_no_compensator}
 P^x_\lambda = \omega^x_u k_\lambda^u \ .
\end{equation}
We review the proof of this in detail in Appendix~\ref{section:prepotential}. Finally, using the explicit form of the vielbein \eqref{quat_vielbein} given above, one can calculate $\omega^x$ in terms of the one-forms \eqref{one-forms_quat} \cite{Ferrara:1989ik}
\begin{equation}\label{quat_connection}
\begin{aligned}
\omega^1\ & =\  \iu (\bar u- u)  \ , \\
\omega^2\ & =\ u + \bar u \ , \\
\omega^3\ & =\ \tfrac{\iu}{2} (v-\bar v) - \iu \e^{K^{\rm h}} \left(Z^A (\Im  \mathcal G_{AB})\diff \bar Z^B - \bar Z^A (\Im  \mathcal G_{AB}) \diff Z^B \right) \ .
 \end{aligned}
\end{equation}

\subsection{Partial Supersymmetry Breaking on Special Quaternionic Manifolds}
\label{breaking_hyper}
Let us now return to the conditions for partial supersymmetry breaking arising from the hypermultiplet sector. The initial analysis in this section follows \cite{Cassani:2007pq}. It will be useful in the following to express the parameter of the unbroken $\cN=1$ supersymmetry in terms of a vector of complex coefficients
\begin{equation} \label{SUSYgenerator}
\epsilon_1^{\cal A}   = \left( \begin{aligned} n^1 \\ n^2 \end{aligned} \right) \epsilon_1 \ ,
\end{equation}
where the Killing spinor $\epsilon_1$ is the generator of the unbroken supersymmetry in $\cN=1$ notation. Inserting \eqref{SUSYgenerator} and \eqref{prepotential_no_compensator} into the gravitino equation \eqref{N=1conditions}, we obtain 
\begin{equation}\label{qS}
 \begin{aligned}
 n^1 u(\kk) + \tfrac{1}{4} n^2 (v-\bar v)(\kk)= \tfrac \iu2
(n^1)^*\,  \e^{-K^{\rm v}/2} \mu \ , \\
 \tfrac{1}{4}n^1 (v-\bar v)(\kk) + n^2 \bar u(\kk) = \tfrac
\iu2 (n^2)^*\,  \e^{-K^{\rm v}/2} \mu  \ ,
 \end{aligned}
\end{equation}
where we have used the following abbreviations for the Killing vectors $k=k^u \partial_u$:
\begin{equation}\label{kVdef}
\kk \equiv {V}^\Lambda \Theta^{\ \lambda}_\Lambda
k_\lambda\ , \qquad \textrm{and} \qquad u(\kk)\equiv \kk^v u_v\ .
\end{equation}
In deriving \eqref{qS}, we also used the fact that the Killing vectors do not have a component in the base directions, i.e.\ $\diff Z^I (k_\lambda) = 0$ holds.

Turning to the hyperino equation \eqref{N=1conditions}, and making use of \eqref{susytrans3}, \eqref{quat_vielbein},
\eqref{one-forms_quat} and \eqref{SUSYgenerator}, we find
\begin{equation}\label{qH}
 \begin{aligned}
n^1 u(\kk) +n^2 v(\kk)  &= 0 \ , \\
  -n^1 \bar v(\kk)+n^2 \bar u(\kk)  &= 0 \ ,
 \end{aligned}
\end{equation}
and
\begin{equation}\label{qE}
\begin{aligned}
  n^2 E^{\underline b}(\kk)  &= 0 \ , \\
  n^1 \bar E^{\underline b}(\kk) &= 0 \ .
\end{aligned}
\end{equation}
In \eqref{qE} we have used that all Killing vectors \eqref{Killing} are in the fibre directions and therefore $e(\kk)=\bar e(\kk)=0$. If we now take the difference of the gravitino \eqref{qS} and hyperino \eqref{qH} conditions, we arrive at
\begin{equation} \label{cmap_vbarv}
\begin{aligned}
n^2 (3v+\bar v) (\kk)  &= - 2 \iu (n^1)^* \e^{-K^{\rm v}/2} \mu  \ , \\
n^1 (v+3\bar v) (\kk)  &= 2 \iu (n^2)^* \e^{-K^{\rm v}/2} \mu  \ .
 \end{aligned}
\end{equation}
Here we see that possible solutions for Minkowski and AdS vacua preserving $\cN=1$  supersymmetry differ significantly due to the $\mu$-term on the right-hand side of \eqref{cmap_vbarv}. By comparing \eqref{cmap_vbarv} with the original hyperino constraint \eqref{qH}, we see that the only way to solve the conditions for a Minkowski vacuum with both $n^1$ and $n^2$ non-zero is to set $v(\kk)=\bar v(\kk)= 0$. As we shall describe further in the next section, one can then easily check that such a vacuum preserves $\cN=2$ supersymmetry \cite{Cassani:2007pq}. Therefore, in order to find an honest $\cN=1$ vacuum we are forced to set $n^1$ or $n^2$ to zero. On the other hand, for AdS vacua  a similar check shows that $n^1$, $n^2$ and $v(\kk)$ must all be non-zero in order to solve \eqref{cmap_vbarv}. Due to the different nature of these possible solutions, we analyse the Minkowski and AdS cases separately in the following.

\subsubsection{Minkowski Vacua}\label{section:Minkowski_vacua}
We will first consider the case of a Minkowski vacuum, setting $\mu=0$ in all the expressions above. As we have just discussed, there are two cases to consider, depending on whether both $n^1$ and $n^2$ are non-zero or not \cite{Frey:2003sd}. If both $n^1$ and $n^2$ are non-zero, one sees from \eqref{cmap_vbarv} that $(v-\bar v)(\kk)=0$ and then the original hyperino conditions \eqref{qH} implies that $u(\kk)=\bar u(\kk)=0$. Inserting this into  \eqref{prepotential_no_compensator} and \eqref{quat_connection} we see that all three prepotentials $P^x$ vanish separately and the vacuum actually has $\cN=2$ supersymmetry \cite{Cassani:2007pq}.\footnote{It is important
to keep in mind that this conclusion crucially depends on the fact that we confine our analysis to the Killing vectors \eqref{Killing}
which correspond to translations in the fibre. If on the other hand isometries in the special K\"ahler base exist, partial supersymmetry might be possible for this case.} If we consider instead the case where one of the components of $n^{\mathcal A}$ is zero we can evade this conclusion. In the remainder of this section we will show that such a solution does exist, and that the conditions for preserved $\cN=1$ supersymmetry \eqref{N=1conditions} can be solved for two commuting isometries.

To proceed, we will set one of the complex coefficients in \eqref{SUSYgenerator} to zero
\begin{equation}\label{nchoice}
n^2=0\ , \qquad n^1\neq 0\ .
\end{equation}
This leads to a simplified set of gravitino \eqref{qS} and hyperino \eqref{qH}, \eqref{qE} equations to solve:
\begin{equation} \label{condition_hypers_M}
v(\kk)=\bar v(\kk)=u(\kk)=\bar E^{\underline b} (\kk) = 0 \ ,
\end{equation}
with $\bar u(\kk)$ and $E^{\underline b}(\kk)$ undetermined. In order to avoid an $\cN=2$ vacuum we must ensure that $\bar u(\kk) \ne 0$, such that $P^x$ does not vanish and we can have the possibility of partial supersymmetry breaking. As we will see, this implies $E^{\underline b}(\kk) \ne 0$.

Our first task is to construct two commuting Killing vectors $k_1$ and $k_2$ out of the set provided by the c-map construction \eqref{isometries_fibre}. By considering the inner product of the quaternionic one-forms \eqref{one-forms_quat} with the Killing vectors \eqref{isometries_fibre}, we see that $\Kdil$ is not a good choice for our purposes as $(v+\bar v) (\Kdil) \ne 0$. Therefore, if we were to use this Killing vector we would not be able to satisfy the $\cN=1$ vacuum conditions \eqref{condition_hypers_M}. This leads us to make the following general ansatz in terms of the remaining Killing vectors
\begin{equation}\label{ex_Killing_vectors}
 \begin{aligned}
  k_1 =  \P_1^B \Kxi_B + \Q_{1\, A} \Ktxi^A + \A_1 \Kax \ , \\
  k_2 =  \P_2^B \Kxi_B + \Q_{2\, A} \Ktxi^A + \A_2 \Kax \ ,
 \end{aligned}
\end{equation}
where for the moment $\P_{1,2}^B, \Q_{1,2\, A}, \A_{1,2}$ are arbitrary real coefficients. By demanding that $k_1$ and $k_2$ commute, we then find a constraint on the coefficients\footnote{At this point, we can already see that we cannot have partial supersymmetry breaking in Minkowski space with just the universal hypermultiplet as the condition \eqref{comm_isometries} reads
\begin{displaymath}
\operatorname{det} \left( \begin{aligned}\P_1 && \P_2 \\ \Q_1 && \Q_2  \end{aligned} \right) = 0 \ .
\end{displaymath}
This in turn means that $k_1$ and $k_2$ are actually linearly dependent, i.e.\ only one linear combination of $\Kxi_A$ and $\Ktxi^A$ is gauged, the prepotentials $P^x_1$ and $P^x_2$ are aligned and no $\cN=1$ solution can be constructed, cf.\ Section~\ref{section:magnetic_vectors}.
}
\begin{equation} \label{comm_isometries}
  \P_1^A \Q_{2\, A} - \P_2^A \Q_{1\, A} = 0 \ .
\end{equation}
If we consider the inner product of the quaternionic one-forms \eqref{one-forms_quat} with our ansatz for the Killing vector combinations \eqref{ex_Killing_vectors}, we immediately observe that both $k_1$ and $k_2$ automatically satisfy the conditions $(v+\bar v) (k_{1,2}) = 0$ , while $(v-\bar v)(\kk) = 0$ imposes
\begin{equation}
{V}^\Lambda \Theta_\Lambda^{\ 1} ( \Q_{1\, A} \xi^A - \P_1^A \tilde\xi_A + \A_1 )  + {V}^\Lambda \Theta_\Lambda^{ \ 2} (\Q_{2\, A} \xi^A - \P_2^A \tilde\xi_A + \A_2) = 0 \ .
\end{equation}
The solution of this condition then fixes the two coefficients $\A_{1}$ and $\A_{2}$
\begin{equation} \label{hypermultiplet_restriction}
 \A_{1,2} = \P_{1,2}^A \tilde\xi_A -\Q_{1,2\, A} \xi^A \ ,
\end{equation}
where $\tilde\xi_A$ and $\xi^A$ are the Ramond-Ramond scalars evaluated at the $N=1$ vacuum. We can now make use of the solution for the embedding tensor components \eqref{solution_embedding_tensor} found from the gravity plus vector multiplet sector, which by construction fulfil \eqref{condition_vectors_Msolve1} and \eqref{condition_vectors_Msolve2}. We already solved the first two equations in \eqref{condition_hypers_M}. Since \eqref{condition_vectors_Msolve1} implies the gravitino and gaugino equation, we find that also $u(\kk)=0$, such that in \eqref{condition_hypers_M} it only remains to solve $\bar E^{\underline b} (\kk) = 0 $, which comes from the hyperino equation and gives further constraints on $\P_{1,2}^A$ and $\Q_{1,2\, A}$. We shall now rewrite the solution for the embedding tensor components \eqref{solution_embedding_tensor} in the notation of this section and then turn to solving the remaining equation $\bar E^{\underline b} (\kk) = 0 $.

Using \eqref{prepotential_no_compensator} and \eqref{quat_connection}, we see that the Killing prepotentials are given by
\begin{equation}
P^+_{1,2} = 2 \iu \bar u(k_{1,2}) \ ,
\end{equation}
where we have used the complex notation introduced in \eqref{Pc}. If we now insert the definition of the one-form $\bar u$ \eqref{one-forms_quat} and make use of \eqref{MG_subspaces}, we find that the solution for the embedding tensor components \eqref{solution_embedding_tensor} can be expressed as
\begin{equation}\label{solution_embedding_tensor_Min}
\begin{aligned}
\Theta_I^{\phantom{I}1} = & -\Im(\bar Z^A (\Q_{2\, A} - \bar {\cal G}_{AB}\P_{2}^B) {\cal F}_{IJ}\C^J ) \ , \\  \Theta^{I1} = & -\Im (\bar Z^A (\Q_{2\, A} - \bar {\cal G}_{AB}\P_{2}^B) \C^I) \ , \\
\Theta_I^{\phantom{I}2} = & \quad \Im(\bar Z^A (\Q_{1\, A} - \bar {\cal G}_{AB}\P_{1}^B) {\cal F}_{IJ} \C^J ) \ , \\  \Theta^{I2} = & \quad \Im (\bar Z^A (\Q_{1\, A} - \bar {\cal G}_{AB}\P_{1}^B)  \C^I) \ ,
\end{aligned}
\end{equation}
where we have absorbed the prefactor $2\e^{K^h/2+\phi}$ into $\C^I$.

Before we solve the condition $\bar E^{\underline b} (\kk) = 0 $, we introduce some techniques from $\cN=2$ supergravity that will prove useful. On any special K\"ahler manifold of dimension $n_{\rm h}-1$ one can define the projection operator ${\Proj}_A^{\phantom{A}B}$ by~\cite{Ferrara:1989ik}
\begin{equation} \label{projection_v}
 {\Proj}_A^{\phantom{A}B} =
\tfrac12 \e^{-K^{\rm h}}{\Proj}_{A\, \underline{b}}
\bar{\Proj}_C^{\phantom{C}\underline{b}} (\Im  \mathcal G)^{-1\,CB} =
\delta_A^B +
2 \e^{K^{\rm h}}{(\Im {\cal G})_{AC}\bar Z^C} Z^B= \delta_A^B + K^{\rm h}_A Z^B\ ,
\end{equation}
where $K^{\rm h}_A$ denotes the holomorphic derivative of the K\"ahler potential $K^{\rm h}$ given in \eqref{Khdef} and ${\Proj}_A^{\phantom{A}\underline{b}}$ is given in \eqref{projector_vielbein}. From the definition follows
\begin{equation}\label{projection_der}
\nabla_a Z^B = \Proj_a^{\phantom{a}B}\ , \qquad \textrm{and} \qquad \nabla_a {\cal G}_B = \Proj_a^{\phantom{a}C}{\cal G}_{CB} \ .
\end{equation}
Furthermore, ${\Proj}_A^{\phantom{A}B}$ has the properties
\begin{equation} \label{properties_projection_v}
Z^A {\Proj}_A^{\phantom{A}B} = 0 \ , \qquad {\Proj}_A^{\phantom{A}B} \Im {\cal G}_{BC} \bar Z^C = 0 \ , \qquad
{\Proj}_A^{\phantom{A}B} {\Proj}_B^{\phantom{B}C} = {\Proj}_A^{\phantom{A}C} \ ,
\end{equation}
and  therefore is indeed a projection map which projects to the space orthogonal to $Z^A$.
From the definition \eqref{projection_v} we we see that ${\Proj}_A^{\phantom{A}B}$ fulfils the reality condition
\begin{equation} \label{relation_projectorbar_v}
 (\Im {\cal G})^{-1 \, DA} \bar{\Proj}_A^{\phantom{A}B} (\Im {\cal G})_{BC} = {\Proj}_C^{\phantom{C}D} \ .
\end{equation}
Furthermore, from \eqref{period_matrix} and \eqref{projection_v} we see that
\begin{equation}\label{MG_subspaces}
Z^A {\cal M}_{AB}=Z^A{\cal G}_{AB} \ , \qquad {\Proj}_A^{\phantom{A}B} \bar{\cal M}_{BC} = {\Proj}_A^{\phantom{A}B} {\cal G}_{BC} \ ,
\end{equation}
which allows us to replace ${\cal M}_{AB}$ in \eqref{one-forms_quat} by ${\cal G}_{AB}$.
The projection ${\Proj}_A^{\phantom{A}B}$ canonically leads to the decompositions
\begin{equation} \label{decomposition}
\begin{aligned}
 \Phi_A =& \Phi^{(Z)}_A + \Phi^{(P)}_A = - K^{\rm h}_A Z^B \Phi_B + {\Proj}_A^{\phantom{A}B} \Phi_B \ , \\
 \Psi^A =& \Psi^{(Z)\,A} + \Psi^{(P)\,A} = - \Psi^B K^{\rm h}_B Z^A  + \Psi^B {\Proj}_B^{\phantom{B}A} \ ,
\end{aligned}
\end{equation}
for any vectors $\Phi_A$ and $\Psi^A$. Note that $\Phi^{(Z)}_A$ and $\Psi^{(Z)\,A}$ each live in a one-dimensional subspace, while
$\Phi^{(P)}_A$ and $\Psi^{(P)\,A}$ parameterise the remaining $n$ directions.
With \eqref{decomposition} we can easily show that $(\Im {\cal G})_{AB}$ is of signature $(n_{\rm h}-1,1)$ \cite{Ceresole:1995ca}:
Using \eqref{MG_subspaces} we find
\begin{equation} \label{signature_G}
\bar \Phi^A (\Im {\cal G})_{AB} \Phi^B= \bar\Phi^{(Z)\,A} (\Im{\cal M})_{AB} \Phi^{(Z)\,B} - \bar\Phi^{(P)\,A} (\Im{\cal M})_{AB} \Phi^{(P)\,B} \ .
\end{equation}
Since $(\Im{\cal M})_{AB}$ is negative definite~\cite{Cremmer:1984hj}, we conclude that $(\Im {\cal G})_{AB}$ is of signature $(n_{\rm h}-1,1)$. 
Note that this result also holds for $(\Im {\cal F})_{IJ}$, with therefore is of signature $(n_{\rm v},1)$.

 
Let us now return to solving $\bar E^{\underline b} (\kk) = 0 $. Inserting \eqref{solution_embedding_tensor_Min} into \eqref{one-forms_quat} we find
\begin{equation}\label{Ebar}
\begin{aligned}
 X^I (\Im {\cal F})_{IJ}  \bar{\C}^J {\Proj}_A^{\phantom{A}B} Z^C \big( & (\Q_{2\, B} - {\cal G}_{BD}\P_{2}^D) (\Q_{1\, C} - {\cal G}_{CE}\P_{1}^E) \\ & - (\Q_{1\, B} - {\cal G}_{BD}\P_{1}^D) (\Q_{2\, C} - {\cal G}_{CE}\P_{2}^E) \big) = 0 \ ,
 \end{aligned}
\end{equation}
where, for convenience, we have contracted the expression with ${\Proj}_{A\, \underline{b}}$ in order to introduce the projection operator $ {\Proj}_A^{\phantom{A}B}$, cf.\ \eqref{projection_v}.
Furthermore, we have used the identity \eqref{identity_v} to pull out the prefactor $ X^I (\Im {\cal F})_{IJ} \bar{\C}^J $. This prefactor is non-zero for all $\C^I$ fulfilling \eqref{constraint_embedding_tensor}, see \eqref{signature_G}, and can be neglected. We can parameterise the Killing vector coefficients $\P_{1,2}^A$ and $\Q_{1,2\, A}$ by
\begin{equation}\label{solution_beta}
 \P_{1,2}^A= \Im(\D^A_{1,2}) \ , \qquad \Q_{1,2\, A} = \Im({\cal G}_{AB}\D^B_{1,2}) \ ,
\end{equation}
where $\D^A_{1,2}$ are two complex vectors. We can then decompose $\D^A_{1,2}$ into the components canonically defined by the projection $ {\Proj}_A^{\phantom{A}B}$ as done in \eqref{decomposition}.
Using this, the condition \eqref{Ebar} simplifies to
\begin{equation}
 \D^{(P)\,A}_{1} \D^{(Z)\,B}_{2} = \D^{(P)\,A}_{2} \D^{(Z)\,B}_{1} \ .
\end{equation}
The only solution to this equation is $\D^A_{2} =a \D^A_{1}$ with a complex factor $a$, and in the following we will just write $\D^A$. Note that for $a$ real, the two Killing vectors are the same and the embedding tensor components \eqref{solution_embedding_tensor_Min} just cancel against each other, giving an ungauged supergravity with an $\cN=2$ vacuum. Furthermore, for any complex $a$, its real part drops out due to this cancellation. Thus, we can choose $a=\iu$, since any additional real prefactor can be absorbed into the embedding tensor. 
After absorbing a prefactor $-\iu \bar Z^A (\Im {\cal G})_{AB} \D^B$ into the definition of $\C^I$, the embedding tensor \eqref{solution_embedding_tensor_Min} similarly to \eqref{solution_embedding_tensor_ex} simply reads
\begin{equation}\label{solution_embedding_tensor_Min2}
\begin{aligned}
\Theta_I^{\phantom{I}1} = & \Im({\cal F}_{IJ}\C^J ) \ , \qquad  \Theta^{I1} = & \Im \C^I \ , \\
\Theta_I^{\phantom{I}2} = & \Re({\cal F}_{IJ} \C^J ) \ , \qquad  \Theta^{I2} = & \Re \C^I \ .
\end{aligned}
\end{equation}
It remains to check that the two Killing vectors commute when the coefficients are parameterised by \eqref{solution_beta}. To do so, we insert \eqref{solution_beta} together with $\D^A= \D^A_{1} = - \iu \D^A_{2}$ into the commutation condition \eqref{comm_isometries} and find
\begin{equation} \label{null_alpha}
 0= \bar \D^A (\Im{\cal G})_{AB} \D^B \ .
\end{equation}
Thus, the complex vector $\D^A$ must be null with respect to the matrix $(\Im{\cal G})_{AB}$, which is of signature $(\nh-1,1)$, cf.\ \eqref{signature_G}.

In order to make contact with the literature, we can rewrite the embedding tensor components in a more convenient basis. Instead of expressing $\Theta_\Lambda^{\ \tilde \lambda}$ in the basis of $k_{1,2}$ plus the other (ungauged) isometries, we can make a change of basis and go back to the standard basis of c-map Killing vectors \eqref{Killing}. To do this, we collect the Killing vectors $\Kxi_A$ and $\Ktxi^A$, as well as the fibre coordinates $\xi^A$ and $\tilde \xi_A$, in the $Sp(\nh)$ vectors
\begin{equation}\label{standard_k}
 k_{\tilde \lambda} = \left( \begin{aligned}
                           \Ktxi^A \\ \Kxi_A
                          \end{aligned} \right) \ ,
\end{equation}
and
\begin{equation} \label{standard_xi}
 \xi_{\tilde \lambda} = \left( \begin{aligned}
                           \xi^A \\ \tilde \xi_A
                          \end{aligned} \right) \ .
\end{equation}
The embedding tensor then reads
\begin{equation} \label{complete_solution_M}
\begin{aligned}
\Theta_\Lambda^{\ \tilde \lambda} &= \Re \left(  \bar{\C}^J \D^B \left( \begin{aligned} \bar {\cal F}_{JI} {\cal G}_{BA} && \bar {\cal F}_{JI} \delta^A_B \\  \delta^I_J {\cal G}_{BA} && \delta^I_J \delta^A_B \end{aligned} \right) \right) \ , \\
\Theta_\Lambda^{\ \ax} &= - \Theta_\Lambda^{\ \tilde \lambda} \xi_{\tilde \lambda} = \Re \left(  \D^A (\tilde\xi_A - {\cal G}_{AB} \xi^B ) \bar{\C}^J \left( \begin{aligned} \bar {\cal F}_{JI} \\  \delta^I_J \end{aligned} \right) \right)  \ ,
\end{aligned}
\end{equation}
where $\D^A$ and $\C^I$ have to satisfy commutation \eqref{null_alpha} and mutual locality \eqref{constraint_embedding_tensor} conditions respectively. 

Before we turn to the AdS case, let us give the explicit form of tensors $S_{\cal AB}$, $W^{i{\cal AB}}$ and $N^\alpha_{\cal A}$ for the embedding tensor solution \eqref{complete_solution_M}:
\begin{subequations}
\begin{eqnarray}
 S_{\cal AB} &=& 2 \e^{K^{\rm v}/2+K^{\rm h}/2+\phi} [X^I (\Im {\cal F})_{IJ} \bar{\C}^J] [\bar Z^A \Im {\cal G}_{AB} \D^B] \left( \begin{aligned}
                    0 && 0 \\ 0 && 1                                                                                  \end{aligned} \right) \ , \\
W_{i \cal AB} &=& 4 \iu \e^{K^{\rm v}/2+K^{\rm h}/2+\phi} [{\Proj}_i^{\phantom{i}J} (\Im {\cal F})_{JK} \bar{\C}^K] [\bar Z^A \Im {\cal G}_{AB} \D^B] \left( \begin{aligned}
                    0 && 0 \\ 0 && 1                                                                                  \end{aligned} \right) \ , \\
N_{\alpha \cal A} &= & 2 \sqrt{2} \iu \e^{K^{\rm v}/2+K^{\rm h}/2+\phi} [X^I (\Im {\cal F})_{IJ} \bar{\C}^J] \hskip4cm \nonumber \\ && \cdot \D^B\left( \begin{aligned}
                    0 && 0 \qquad && 0 \qquad && 0\\ 0 && [\tfrac12 \e^{-K^{\rm h}} {\Proj}_B^{\phantom{B}\underline{a}}] && [(\Im {\cal G})_{BA} \bar Z^A] && 0   \end{aligned} \right) \ ,
\end{eqnarray}
\end{subequations}
where we used the relations between the projector $\Proj_i^{\phantom{i}J}$ and the K\"ahler covariant derivatives of $X^J$ and ${\cal F}_J$ \eqref{projection_der}.

Note that the solution \eqref{complete_solution_M} can be constructed for \emph{any} point of the moduli space $\M_{\rm v}\times \M_{\rm h}$ and does only depend on the second derivatives of the prepotentials $\cal F$ and $\cal G$ at the $\cN=1$ point.
Furthermore, the solution is completely covariant under Mirror symmetry, which essentially exchanges the two special K\"ahler manifolds.

\subsubsection{AdS Vacua}\label{section:AdS_vacua}

Let us now consider the case of an AdS vacuum preserving $\cN=1$ supersymmetry. For $\mu \ne 0$, we see from combined gravitino and hyperino condition \eqref{cmap_vbarv} that both $n^1$ and $n^2$ must be non-zero. By manipulating \eqref{cmap_vbarv}, we are led to the following conditions
\begin{subequations}\label{AdS_vbarv}
 \begin{eqnarray}
  n^1 n^2 (v+\bar v) (\kk) &=& - \tfrac12 \iu \e^{-K^{\rm v}/2}\mu (|n^1|^2 - |n^2|^2) \ , \label{AdS_vbarv1} \\
  n^1 n^2 (v-\bar v) (\kk) &=& - \iu \e^{-K^{\rm v}/2}\mu (|n^1|^2 + |n^2|^2) = - \iu \e^{-K^{\rm v}/2}\mu |\epsilon_1|^2\ . \label{AdS_vbarv2}
 \end{eqnarray}
\end{subequations}
If the $k_\phi$ direction is not gauged, then we have that $(v+\bar v) (k(L))=0$ and we can conclude that the complex coefficients of the preserved supersymmetry generator \eqref{SUSYgenerator} must be equal $|n^1|= |n^2|$ \cite{Cassani:2007pq}.\footnote{The dilaton isometry is spoilt by quantum corrections in $\cN=2$ supergravity. Therefore we do not consider gaugings with respect to this isometry.} This agrees with the result using a different approach in type II supergravity in ten dimensions \cite{Grana:2006kf}. In the following we shall restrict to $|n^1|= |n^2|$ and parameterise the coefficients as
\begin{equation}  \label{unbroken_SUSY_AdS}
n^1 =\e^{\iu \varphi/2}n \ , \quad \textrm{and} \quad n^2=\e^{-\iu \varphi/2}n \ ,
\end{equation}
where $\varphi$ is a phase.

Before we proceed to analyse the supersymmetry variations in detail, we shall make a remark about the amount of unbroken supersymmetry. For AdS vacua, we take the general ansatz for the Killing vectors $k_1$ and $k_2$  used in the Minkowski case \eqref{ex_Killing_vectors}, and demand that they commute i.e. that \eqref{comm_isometries} is satisfied. The embedding tensor components which solve the gravitino and gaugino equations are then given by \eqref{solution_embedding_tensor_AdS}, but as we now break to a different $\cN=1$ vacuum with a different preserved Killing spinor \eqref{SUSYgenerator} we must perform an $SU(2)$-rotation. By comparing \eqref{unbroken_SUSY_AdS} with the spinor used in Section \eqref{section:magnetic_vectors}, which has $n^1\ne 0$ and $n^2=0$, we see that the appropriate $SU(2)$-rotation is given by
\begin{equation}
 M^{\cal A}_{\phantom{\cal A} \cal B}= \tfrac{1}{\sqrt{2}}\left( \begin{aligned}
           \e^{\iu\varphi/2} && - \e^{\iu\varphi/2} \\ \e^{-\iu\varphi/2} &&\e^{-\iu\varphi/2}
           \end{aligned}\right) \ .
\end{equation}
The only term in the embedding tensor components \eqref{solution_embedding_tensor_AdS} that transforms non-trivially under this rotation is $P^+_{1,2}$:
\begin{equation} \label{twisted_prepotentials}
 P^-_{1,2} \longrightarrow \tilde P^-_{1,2} = \iu \Im(\e^{\iu \varphi} P^-_{1,2}) - P^3_{1,2} \ .
\end{equation}
In order to find the embedding tensor components which solve the gravitino and gaugino conditions \eqref{solution_embedding_tensor_AdS} we assumed that $P^3_{1,2}=0$. In the new $SU(2)$-frame we have to adjust $k_1$ and $k_2$ such that
\begin{equation} \label{vanishingP3}
 \tilde P^3_{1,2} = \Re(\e^{\iu \varphi} P^-_{1,2}) = 0 \ .
\end{equation}
Analogously to the Minkowski case \eqref{solution_beta}, we make the following ansatz for the Killing vector coefficients
\begin{equation}\label{solution_beta_AdS}
 \P_{1,2}^A= \Im(\D^A_{{\rm AdS}\,1,2}) \ , \qquad \Q_{1,2\, A} = \Im({\cal G}_{AB}\D^B_{{\rm AdS}\,1,2}) \ ,
\end{equation}
where we have used the decomposition \eqref{decomposition} with respect to the projector ${\Proj}_A^{\phantom{A}B}$ to express $\D^A_{{\rm AdS}\,1,2}$ as
\begin{equation} \label{decompose_AdS}
 \D^A_{{\rm AdS}\,1,2} =  \D^{(Z)\,A}_{{\rm AdS\,}1,2} + \D^{(P)\,A}_{{\rm AdS}\,1,2} \ .
\end{equation}
Inserting this ansatz into \eqref{vanishingP3} and using the expressions \eqref{prepotential_no_compensator}, \eqref{quat_connection} and \eqref{one-forms_quat} we find
\begin{equation}
 \Re(\e^{\iu \varphi} Z^A (\Im {\cal G})_{AB} \bar{\D}^{(Z)\,B}_{{\rm AdS}\,1,2}) = 0 \ ,
\end{equation}
which is solved by
\begin{equation} \label{TZ}
 \D^{(Z)\,A}_{{\rm AdS\,}1,2} = \iu \e^{\iu \varphi} \R_{1,2} Z^A\ ,
\end{equation}
where $\R_{1,2}$ are real numbers.
Inserting the above expressions into the transformation of the Killing prepotential \eqref{twisted_prepotentials} then leads to
\begin{equation} \label{prepotentials_AdS}
 \tilde P^-_{1,2} = \e^{2\phi} (\A_{1,2} -  \Im((\iu \R_{1,2} \e^{\iu\varphi}Z^A +\bar{\D}^{(P)\,A}_{{\rm AdS}\,1,2}) (\tilde \xi_A - {\cal G}_{AB} \xi^B)) ) + \iu \e^{-K^{\rm h}/2 + \phi} \R_{1,2} \ .
\end{equation}
We remind the reader that the prepotentials $\tilde P^x_1$ and $\tilde P^x_2$ should not be aligned for a proper $\cN=1$ vacuum.

We still have to solve the equations coming from the hyperino variation. In the Minkowski case we only had to solve the condition $\bar E(\kk)=0$, whereas we now see from \eqref{qE} that we that we have an addition condition $E(\kk)=0$ in the AdS case. Furthermore, \eqref{qH} also now gives an additional non-trivial condition, which is rephrased as \eqref{AdS_vbarv2}. Considering again the projector decomposition \eqref{decomposition} for $\D_{{\rm AdS}\,1,2}^A$,  we see that \eqref{AdS_vbarv2} gives a condition on $\C_{1,2}$, while \eqref{qE} restricts $\D^{(P)\,A}_{{\rm AdS}\,1,2}$ in \eqref{decompose_AdS}. Let us start with \eqref{qE}. By plugging in \eqref{ex_Killing_vectors} with \eqref{solution_beta_AdS} and using the definition \eqref{projection_v} and the relations \eqref{MG_subspaces}, we can write \eqref{qE} as
\begin{equation}\label{AdS_qE}
 \begin{aligned}
 (\tilde P^2_2 + \tfrac\iu2 \tilde P^1_2) \D^{(P)\,A}_{{\rm AdS}\,1}  - (\tilde P^2_1 + \tfrac\iu2 \tilde P^1_1)\D^{(P)\,A}_{{\rm AdS}\,2} = 0     \ , \\
 (\tilde P^2_2 - \tfrac\iu2 \tilde P^1_2) \D^{(P)\,A}_{{\rm AdS}\,1}-  (\tilde P^2_1 - \tfrac\iu2 \tilde P^1_1)\D^{(P)\,A}_{{\rm AdS}\,2} = 0 \ ,
 \end{aligned}
\end{equation}
where for simplicity we took  the complex conjugate in the first equation. As the prepotentials of $k_1$ and $k_2$ must not coincide in an $\cN=1$ vacuum, \eqref{AdS_qE} implies that both $\D^{(P)\,A}_{{\rm AdS}\,1}$ and $\D^{(P)\,A}_{{\rm AdS}\,2}$ must vanish. Then from the commutation relation \eqref{comm_isometries}, together with \eqref{solution_beta_AdS}, \eqref{decompose_AdS} and \eqref{TZ}, it follows that $\R_1$ or $\R_2$ is zero. We can choose $\R_2=0$ and note that by taking linear combinations of $k_1$ and $k_2$ we can always set $\A_1=0$. Furthermore, the resulting Killing vectors can be rescaled such that $\R_1=\A_2=1$.

Let us now solve \eqref{AdS_vbarv2}. Inserting the embedding tensor \eqref{solution_embedding_tensor_AdS} with \eqref{prepotentials_AdS} and \eqref{solution_beta_AdS}, we find
\begin{equation}\label{AdS_fixing}
 X^I (\Im {\cal F})_{IJ} \bar{\C}_{\rm AdS}^J = \iu \frac{3+4 \iu \E}{2+2 \iu \E} \e^{K^{\rm h}/2-K^{\rm v}/2-3\phi} \mu  \ ,
\end{equation}
where we abbreviated
\begin{equation}
  \E = \e^{K^{\rm h}/2 +\phi} \Re(\e^{\iu \varphi} (Z^A \tilde \xi_A - {\cal G}_A\xi^A)) \ .
\end{equation}
Using again the decomposition \eqref{decomposition} we can insert \eqref{AdS_fixing} into \eqref{solution_embedding_tensor_AdS}.
If we now go back to the standard basis of \eqref{standard_k} and \eqref{standard_xi}, the embedding tensor reads
\begin{equation}\label{complete_solution_AdS}
\begin{aligned}
\Theta_\Lambda^{\ \tilde \lambda} &= - \Re \left( \left( \begin{aligned} {\cal F}_{IJ} \\  \delta^I_J \end{aligned} \right) (4\e^{K^{\rm h}/2 + K^{\rm v}/2-\phi} \bar \mu X^J + \C^{(P)\,J}_{\rm AdS})  \right) \cdot \Re( \e^{\iu \varphi}(\ {\cal G}_{A}  \ ,\  Z^A \ ))  \ , \\
\Theta_\Lambda^{\ \ax} &= \e^{-K^{\rm h}/2 -\phi} \Im \left( \left( \begin{aligned} {\cal F}_{IJ} \\  \delta^I_J \end{aligned} \right) (4 \e^{K^{\rm h}/2 + K^{\rm v}/2-\phi} (\tfrac12- \iu \E) \bar \mu X^J + (1-\iu \E) \C^{(P)\,J}_{\rm AdS})  \right)  \ ,
\end{aligned}
\end{equation}
where we have rescaled $\C^{(P)\,I}_{\rm AdS}$ by the factor $\iu\e^{2\phi}$. If we plug our result \eqref{complete_solution_AdS} into the constraint \eqref{constraint_embedding_tensor_AdS_g}, we find
\begin{equation}\label{constraint_AdS}
 \bar \C^{(P)\,J}_{\rm AdS} (\Im {\cal F})_{JI} \C^{(P)\,I}_{\rm AdS} = \e^{K^{\rm h} - 6\phi} \frac{|\mu|^2}{1+\rho^2} \ .
\end{equation}
This can be easily solved, since the left-hand side is naturally greater than zero (see the discussion in \eqref{signature_G}). The solution \eqref{complete_solution_AdS} should correspond to the result of \cite{Cassani:2009na}.

Finally, for the embedding tensor solution \eqref{complete_solution_AdS} the tensors appearing in the supersymmetry transformations $S_{\cal AB}$, $W^{i{\cal AB}}$ and $N^\alpha_{\cal A}$ are given by
\begin{subequations}
\begin{eqnarray}
 S_{\cal AB} &=& \mu  \left( \begin{aligned}
                    \e^{-\iu \varphi} && -\tfrac12 \\ -\tfrac12 && \e^{\iu \varphi}                                                                                  \end{aligned} \right) \ , \hskip4.5cm \\
W_{i \cal AB} &=& -\tfrac12 \e^{K^{\rm v}/2-K^{\rm h}/2+\phi} (\Im {\cal F})_{iJ} \bar \C^{(P)\,J}_{\rm AdS}
               \left( \begin{aligned}
                    \e^{-\iu \varphi} && -1 \\ -1 && \e^{\iu \varphi}                                                \end{aligned} \right) \ , \\
N_{\alpha \cal A} &= & \tfrac{1}{\sqrt{2}} \iu \mu \left( \begin{aligned}
                    \e^{-\iu \varphi} && 0 && -1 && 0\\ -1 && 0 && \e^{\iu \varphi} && 0   \end{aligned} \right) \ ,\hskip3.5cm
\end{eqnarray}
\end{subequations}
where we have again used \eqref{projection_der}. 

The embedding tensor given by \eqref{complete_solution_AdS} can be defined at any point on ${\M}_{\rm v}\times {\M}_{\rm h}$. Furthermore, for any choice of the moduli spaces ${\M}_{\rm v}$ and ${\M}_{\rm h}$ -- as long as ${\M}_{\rm h}$ is in the image of the c-map -- we have found a construction for the gaugings that lead to $\cN=1$ AdS vacua. The only constraints on the solution \eqref{complete_solution_AdS} is \eqref{constraint_AdS}, which can easily be fulfilled. In this way, the results of this section are completely analogous to those of Section~\ref{section:Minkowski_vacua}.  

\section{Realisation in String Theory} \label{section:strings}
Let us now show how the solutions of Section~\ref{section:hypermultiplets} can be realised in string theory. We shall only consider $\cN=2$ compactifications of the type II string here, but similar realisations should be possible for the heterotic string. For notational simplicity we restrict our discussion to type IIA. The type IIB results are then easily obtained by exchanging even and odd forms. For further discussion of four-dimensional $\cN=1$ Minkowski and AdS vacua from string theory see \cite{Lust:2004ig,Behrndt:2005bv,Grana:2005sn,Grana:2006kf,Micu:2007rd,KashaniPoor:2007tr,Andriot:2008va,Anguelova:2008fm,Cassani:2009ck,Lust:2009zb}.

The  $\cN=2$ string compactifications that we consider in the following have an internal, six-dimensional manifold $Y$  which admits an
$SU(3)\times SU(3)$-structure, see for instance \cite{Gurrieri:2002wz,Grana:2005ny,Grana:2006hr,Cassani:2008rb}. The light modes are  obtained from the ten-dimensional fields by expanding in a finite-dimensional symplectic basis of even forms $\omega_I$, $\tilde \omega^I$ and a finite-dimensional symplectic basis of odd forms $\alpha_A$ and $\beta^A$. None of these forms are necessarily closed, but rather they obey
\begin{equation} \label{forms}
\begin{aligned}
\diff \alpha_A = & p_A^I \omega_I + e_{AI} \tilde \omega^I \ , \qquad \quad
\diff \beta^A = & q^{AI} \omega_I + m^A_I \tilde \omega^I \ , \\
\diff \omega_I = & m^A_I \alpha_A - e_{AI} \beta^A \ , \qquad \
\diff \tilde \omega^I = & - q^{AI} \alpha_A - p_A^I \beta^A  \ ,
\end{aligned}
\end{equation}
where $e_{AI}, m^A_I, p_A^I, q^{AI}$ are constant matrices parameterising the intrinsic torsion of $Y$ as well as background flux
of the NS three-form $H$. The parameters $e_{AI}$ and $m^A_I$ already appear in $SU(3)$-structure compactification while $p_A^I$ and $q^{AI}$ only arise in genuine $SU(3)\times SU(3)$-structure compactifications and are often referred to as non-geometric fluxes. Additionally, there can be background flux for the Ramond-Ramond form fields $F = F_0 + F_2 + F_4 + F_6 = \e^B\wedge G$, which is expanded as
\begin{equation}
 G = \sqrt{2}\, (m^I_{\rm RR} \omega_I + e_{{\rm RR} \, I} \tilde \omega^I ) \ .
\end{equation}

Refs.\ \cite{Grana:2005ny,Grana:2006hr} determined the gravitino mass matrix $S_{\cal AB}$ for this class of compactifications. By comparing their result with \eqref{susytrans3} we can read off the corresponding embedding tensor as
\begin{equation}\label{Etensorcharges}
 \Theta^{\ \tilde \lambda}_\Lambda = \left( \begin{aligned}
                            e_{AI} && p_A^I \\ m^A_I && q^{AI}
                          \end{aligned} \right) \ , \qquad
\Theta^{\ \ax}_\Lambda = (e_{{\rm RR} \, I} \ , \ m^I_{\rm RR} ) \ .
\end{equation}
$\Theta^{\tilde \lambda}_\Lambda$ precisely coincides with the `doubly symplectic' charge matrix ${\cal Q}$ discussed in \cite{Berglund:2005dm,Dall'Agata:2006nr,Grana:2006hr}. Note that the locality constraint \eqref{locality_two_isometries} and the commutativity of the two Killing vectors \eqref{comm_isometries} form the quadratic constraints of ${\cal Q}$, as discussed in \cite{D'Auria:2004wd,D'Auria:2007ay}.

For the $\cN=1$ Minkowski solution \eqref{complete_solution_M} we can identify the charges appearing in \eqref{Etensorcharges} as follows
\begin{subequations}\label{Msolution}
\begin{eqnarray}
e_{AI} &= & \Re(\bar {\cal F}_{IJ} \bar{\C}^J {\cal G}_{AB} \D^B)  \ , \\
p_A^I &= & \Re(\bar{\C}^I {\cal G}_{AB} \D^B) \ , \\
m^A_I  &= & \Re(\bar {\cal F}_{IJ} \bar{\C}^J \D^A) \ ,\\
q^{AI}  &= & \Re(\bar{\C}^I \D^A) \ ,\\
e_{{\rm RR}\,I}  &= & \Re(\bar {\cal F}_{IJ} \bar{\C}^J (\xi^A {\cal G}_{AB}- \tilde \xi_B )\D^B) \ , \\
m^I_{\rm RR} &= & \Re(\bar{\C}^I (\xi^A {\cal G}_{AB}- \tilde \xi_B )\D^B) \ .
\end{eqnarray}
\end{subequations}
Let us recall that charges are quantised in string theory and therefore all entries of the embedding tensor are integral. This implies that partial supersymmetry breaking may only be possible at discrete points on $\M_{\rm v}$ and $\M_{\rm h}$, where the expressions in \eqref{Msolution} are integer-valued. This condition might restrict the form of the prepotential and therefore the allowed moduli spaces $\M_{\rm v} \times \M_{\rm h}$.

The issue of mirror symmetry in $SU(3)\times SU(3)$-structure compactifications has been discussed at length in Ref.\ \cite{Grana:2006hr}, where it was found that, apart from an exchange of the prepotentials ${\cal F} \leftrightarrow {\cal G}$, the charges are exchanged as follows \begin{equation}
m^A_I \leftrightarrow -p^A_I\ , \qquad e_{AI}\leftrightarrow
e_{IA}\ ,\qquad q^{AI}\leftrightarrow q^{IA}\ .
\end{equation}
An inspection of \eqref{Msolution} shows that the solutions indeed obey this symmetry if we also simultaneously exchange $\C^I\leftrightarrow  \D^A$.

If we set $p_A^I$ and $q^{AI}$ to zero in \eqref{Msolution}, the product $\bar \C^I \D^B$ must vanish and we end up with the trivial solution. Therefore, an $\cN=1$ Minkowski vacuum can only occur when non-geometric fluxes are turned on. This is in agreement with the compactification no-go-theorem \cite{Gibbons:1984kp,deWit:1986xg,Maldacena:2000mw}, which states that there can be no stable Minkowski vacuum with only fluxes turned on. This statement is believed to also be true for backgrounds with torsion. Here we explicitly see that non-geometric fluxes can compensate for the form field fluxes and torsion, leading to a vanishing energy density i.e.\ to vanishing $\mu$. In this way, the solution of Section~\ref{section:Minkowski_vacua} evades the no-go theorem.\footnote{A related result on the necessity of non-geometric fluxes for Minkowski vacua in orientifold compactifications has recently been found \cite{deCarlos:2009qm}.}  

Before we turn to the AdS case, let us also note that the $\cN=1$ solutions given in \eqref{Msolution} are not within the class of solutions considered in \cite{Grana:2005sn} as one of the complex parameters $n^1$ or $n^2$ introduced in \eqref{SUSYgenerator} has to vanish. Rather, they  correspond to the class of solutions denoted Type A in \cite{Frey:2003sd}, which have been much less investigated. It would be interesting to further investigate this class of models.

We shall now consider the solution for $\cN=1$ AdS vacua. Comparing \eqref{complete_solution_AdS} with \eqref{Etensorcharges} we can read off
\begin{subequations}\label{Asolution}
\begin{eqnarray}
e_{AI} &= & -\Re( {\cal F}_{IJ} (4\e^{K^{\rm h}/2 + K^{\rm v}/2-\phi} \bar \mu X^J + \C^{(P)\,J}_{\rm AdS}) ) \quad \Re( \e^{\iu \varphi}  {\cal G}_{A} ) \ , \\
p_A^I &= & -\Re( (4\e^{K^{\rm h}/2 + K^{\rm v}/2-\phi} \bar \mu X^I + \C^{(P)\,I}_{\rm AdS}) \quad \Re( \e^{\iu \varphi}  {\cal G}_{A} ) \ , \\
m^A_I  &= & -\Re( {\cal F}_{IJ} (4\e^{K^{\rm h}/2 + K^{\rm v}/2-\phi} \bar \mu X^J + \C^{(P)\,J}_{\rm AdS}) ) \quad \Re( \e^{\iu \varphi} Z^A) \ ,\\
q^{AI}  &= & -\Re( (4\e^{K^{\rm h}/2 + K^{\rm v}/2-\phi} \bar \mu X^I + \C^{(P)\,I}_{\rm AdS}) \quad \Re( \e^{\iu \varphi} Z^A) \ ,\\
e_{{\rm RR}\,I}  &= & \e^{-K^{\rm h}/2 -\phi}\Im( {\cal F}_{IJ} (4 \e^{K^{\rm h}/2 + K^{\rm v}/2-\phi} (\tfrac12- \iu \E) \bar \mu X^J + (1-\iu \E) \C^{(P)\,J}_{\rm AdS}) ) \ , \\
m^I_{\rm RR} & = & \e^{-K^{\rm h}/2 -\phi} \Im( 4 \e^{K^{\rm h}/2 + K^{\rm v}/2-\phi} (\tfrac12- \iu \E) \bar \mu X^I + (1-\iu \E) \C^{(P)\,I}_{\rm AdS} ) \ .
\end{eqnarray}
\end{subequations}
If we turn off non-geometric fluxes ($p_A^I=q^{AI} =0$), we see that non-trivial solutions do exist but must obey
\begin{equation}
\Re (X^I \bar \mu) = 0 \ .
\end{equation}
It would be interesting to further investigate the ten-dimensional origin of this condition. 

Let us close this section by discussing possible quantum corrections in string theory. First of all, worldsheet instantons correct the K\"ahler potentials $K^{\rm v}$ in type IIA and $K^{\rm h}$ in type IIB. However, since we never used their explicit forms, all our results are unchanged and hold for any instanton-corrected K\"ahler potential. What we did use explicitly were the isometries resulting from the special fibration structure of $\M_{\rm h}$. Spacetime instanton effects generated from wrapped Euclidean branes generically break all of the isometries of $\M_{\rm h}$. However, it has been argued that the isometries which are gauged due to fluxes are precisely those protected (by the flux itself) from spacetime instanton effects \cite{KashaniPoor:2005si}.
It would be very interesting to identify \eqref{Msolution} and \eqref{Asolution} as solutions of the ten-dimensional supergravity equations of motion.

\section{Conclusions} \label{section:conclusions}

We have carried out a systematic analysis of when spontaneous $\cN=2 \rightarrow \cN=1$ supersymmetry breaking can take place in gauged supergravities with general vector multiplet couplings and special hypermultiplet couplings. Our results provide a new perspective on the circumvention of well-known no-go theorems which forbid partial supersymmetry breaking in a Minkowski vacuum for a class of supergravity theories \cite{Cecotti:1984rk,Cecotti:1984wn,Mayr:2000hh,Maldacena:2000mw}. In particular, we have found the general solution to the conditions for spontaneous $\cN=2 \rightarrow \cN=1$ supersymmetry breaking in Minkowski and AdS space. 

In contrast to the known examples in the literature \cite{Ferrara:1995gu,Ferrara:1995xi,Fre:1996js}, we have worked directly in a rotated symplectic frame in which a holomorphic prepotential $\cF$ exists and mutually local electric and magnetic charges are introduced. By considering the symplectic extension of the $\cN=2$ supersymmetry variations and initially focussing on the gravitino and gaugino equations, we were able to derive a set of conditions for spontaneous partial supersymmetry breaking in terms of the charges, encoded in the embedding tensor. We then derived the general solution to these conditions by assuming the existence of an appropriate pair of commuting Killing vectors. For the Minkowski case the solution is such that in the purely electric frame the prepotential does not exist at the $\cN=1$ point. Furthermore, the conditions are insensitive to the explicit form of the K\"ahler potential $K^{\rm v}$ and thus to any quantum corrections to the prepotential $\cF$ e.g.\ due to worldsheet instantons. This led us to  conclude that solving the conditions for spontaneous $\cN=2 \rightarrow \cN=1$ supersymmetry breaking in Minkowski or AdS vacua imposes conditions on the charges of the theory (i.e.\ the embedding tensor components), but not on the special K\"ahler geometry.

To complete our analysis, we then turned to the constraints arising from the hyperino variations. By focussing on the case of special quaternionic-K\"ahler manifolds, we could construct two commuting Killing vectors out of the Heisenberg algebra of Killing vectors that arises in the c-map construction such that they solve the additional necessary conditions coming from the hyperino variation.
The resulting solutions for the embedding tensor components could be rephrased in terms of the second derivatives of the prepotentials. For the Minkowski case, we found that the set of conditions for partial supersymmetry breaking are mirror symmetric under the exchange of the prepotentials of the special K\"ahler ($\cF$) and special quaternionic-K\"ahler ($\cG$) geometry. By considering how the parameter $\epsilon_1$ of the preserved $\cN=1$ supersymmetry in a Minkowski vacuum is related to the original pair of $\cN=2$ parameters, we also found that the solutions lie outside of those usually considered in the pure spinor approach to flux compactifications \cite{Grana:2006kf}, as one of the complex coefficients of the spinors has to vanish. Rather, they are the Type A  vacua in the classification scheme described in \cite{Grana:2005jc}. For an $\cN=1$ AdS vacuum, on the other hand, we found that absolute value of the spinor coefficients had to be equal, in agreement with the result derived from ten dimensions \cite{Grana:2006kf}. Our final conclusion is that spontaneous $\cN=2 \rightarrow \cN=1$ supersymmetry breaking is possible at any point on the special K\"ahler manifold and at any point on the special quaternionic-K\"ahler manifold in gauged supergravity.

It would be useful to derive the low-energy effective theory arising after spontaneous $\cN=2 \rightarrow \cN=1$ supersymmetry breaking. Of particular interest for moduli stabilisation is the question of which masses are generated by the partial supersymmetry breaking and to what extent is it possible to find chiral $\cN=1$ theories. Some initial results in this direction appear in Appendix~\ref{section:stability}, where we show that the $\cN=1$ vacua found here are stable by analysing the derivatives of the scalar potential and derive the mass term for the scalars in terms of the mass matrices of the spin 1/2 particles.  It would also be interesting to understand how to extend our analysis to more general quaternionic-K\"ahler manifolds, outside of the special class considered here, and in particular what are the requirements for isometries on general quaternionic-K\"ahler manifolds.

It is natural to ask about the stringy realisation of this mechanism for partial supersymmetry breaking. By comparing our solution for the embedding tensor components with the charges appearing in flux compactifications, we found that the charges needed to solve the $\cN=1$ Minkowski vacuum conditions include non-geometric fluxes. This explains how we have evaded the no-go theorem forbidding the compactification of supergravity to Minkowski space in four dimensions \cite{Gibbons:1984kp,deWit:1986xg,Maldacena:2000mw}, which applies only to geometric fluxes. For an $\cN=1$ AdS vacuum, we found that geometric fluxes alone are sufficient to solve the supersymmetry conditions. For both cases, a possible direction for future work would be to understand the lift of the general $\cN=1$ solutions.

Finally, we should note that the fluxes appearing in a supergravity derived from string theory are quantised, and therefore partial supersymmetry breaking may only be possible at discrete points on $\M_{\rm v} \times \M_{\rm h}$, where the second derivatives of the prepotentials obey an integer condition. Furthermore, flux quantisation may put some constraints on the allowed moduli spaces. We shall leave a more thorough analysis of this point for future work.

\vskip 1cm

\subsection*{Acknowledgements}

This work was supported by the German Science Foundation (DFG) under the Collaborative Research Center (SFB) 676. We have greatly benefited from conversations and correspondence with Davide Cassani, Vicente Cortes, Bernard de Wit, Bobby Gunara, Dieter L\"ust, Peter Mayr, Michela Petrini, Henning Samtleben, Stefan Vandoren and Antoine Van Proeyen.

\vskip 1cm

\appendix
\noindent
{\bf\Large Appendix}

\section{Conventions and Technical Details}\label{appA}

\subsection{$SU(2)$ Matrices}

The $SU(2)$ matrices $(\sigma^x)_{\cal AB}$ which appear in the $\cN=2$ supersymmetry variations are given by
\begin{equation}
(\sigma^1)_{\cal AB} =  \left(\begin{array}{cc}1 &0\\ 0& -1 \end{array}\right)~, \qquad (\sigma^2)_{\cal AB} =  \left(\begin{array}{cc} -\iu &0\\ 0& -\iu \end{array}\right)~ , \qquad (\sigma^3)_{\cal AB} =  \left(\begin{array}{cc}0 &-1\\ -1& 0 \end{array}\right)~.
\end{equation}
These can be found from the usual Pauli matrices by applying the antisymmetric $SU(2)$ metric $\epsilon_{AB}$, which in our conventions has the properties
\begin{equation}
\epsilon^{\cal AB} \epsilon_{\cal BC}  = - \delta^{\cal A}_{\cal C}~, \qquad \epsilon^{12}  = \epsilon_{12} =+1~.
\end{equation}

\subsection{Vector Multiplets Coupled to $\cN=2$ Supergravity}
\label{section:sg}
In this appendix we supplement our discussion of $\cN=2$ gauged supergravity in $D=4$ in Section~\ref{section:N=2} with some further details. For a comprehensive review see e.g.\ \cite{Andrianopoli:1996cm}. 

$\cN=2$ supergravity coupled to $\nv$ vector multiplets contains $\nv+1$ gauge bosons $A_\mu^I, I=0,\ldots, \nv$ together with  $\nv$ complex scalars $t^i, i = 1,\ldots,\nv $ as bosonic components. In the ungauged case the Lagrangian reads
\begin{equation}\label{Lvect}
{\cal L}\ =\ - \mathrm{Im} \mathcal{N}_{IJ}\,
F^{I}_{\mu\nu}F^{\mu\nu\, J} - \mathrm{Re} \mathcal{N}_{IJ}\,
F^{I}_{\mu\nu} F_{\rho\sigma}^{J}\epsilon^{\mu\nu\rho\sigma}
+ g_{i\bar \jmath}\, \partial_\mu t^i \partial_\mu\bar
t^{\bar \jmath} \ ,
\end{equation}
where $F^{I}=dA^I$ are the Abelian field strengths of the $A^I$. The $t^i$ span a special K\"ahler manifold $\M_{\rm v}$, i.e.\ the K\"ahler potential $K^{\rm v}$ is determined by the two holomorphic vectors $(X^I(t), \cF_I(t))$ to be $K^{\rm v}= -\ln \iu(\bar X^I \cF_I - X^I\bar \cF_I)$. The matrix of gauge couplings is also expressed in terms of these vectors: 
\begin{equation}
  \label{Ndef}
  {\cal N}_{IJ} = \bar \cF_{IJ} +2\iu\ \frac{\mbox{Im} \cF_{IK}\mbox{Im}
    \cF_{JL} X^K X^L}{\mbox{Im} \cF_{LK}  X^K X^L} \ ,
\end{equation}
where $\cF_{IJ}=\partial_I \cF_J$.

The equations of motion derived from the action \eqref{Lvect} are
invariant under
generalised symplectic $Sp(\nv+1)$ electric-magnetic duality
transformations.
They act on the $(2\nv+2)$-dimensional
symplectic vector $H^\Lambda\equiv (F^I, G_I)$ according to
\begin{equation}\label{Sptrans}
H^\Lambda \to H^{\prime \Lambda} = {{\cal S}^\Lambda}_\Sigma H^\Sigma\ ,
\end{equation}
where $G_I\equiv \partial L/\partial F^I$ is the field
strength of the dual magnetic gauge boson. ${\cal S}$ is an
$(2\nv+2)\times(2\nv+2)$ matrix 
which leaves the metric
$\Omega$ of $Sp(\nv+1)$ invariant, i.e.\ $S$
obeys ${\cal S}\Omega {\cal S}=\Omega$, where the metric $\Omega$ is given by
\begin{equation}
 \Omega = \left( \begin{aligned}
                  0 && \mathbbm{1}  \\ -\mathbbm{1} &&  0
                \end{aligned} \right) \ .
\end{equation}
In terms of $(\nv+1)\times(\nv+1)$ matrices $S$ is given by
\begin{equation}
  \label{uvzwg}
  {\cal S}\ = \left(
    \begin{array}{cc}
      U & Z \\
      [1mm] W & V
    \end{array}
  \right) \ ,
\end{equation}
where $U$, $V$, $W$ and $Z$ 
obey
\begin{equation}\begin{aligned}
  \label{spc2}
  U^{\rm T} V- W^{\rm T} Z &= V^{\rm T}U - Z^{\rm T}W =
  {\bf 1}\, ,\\
  U^{\rm T}W = W^{\rm T}U\,, & \quad Z^{\rm T}V= V^{\rm T}Z\ .
\end{aligned}
\end{equation}

${V}^\Lambda=(X^I,\cF_I)$ is a symplectic vector and transforms according to \eqref{Sptrans}. The K\"ahler potential is invariant under symplectic rotations, as can be easily seen by rewriting it in a symplectic invariant form
\begin{equation}
  \label{Ksymp}
 K^{\rm v}= -\ln \iu \left(\bar {V}^\Lambda \Omega_{\Lambda\Sigma} {V}^\Sigma \right) \ .
\end{equation}
The kinetic matrix $\cN$ on the other hand transforms according to
\begin{equation}
  \label{nchange}
  \cN \to (V \cN+ W) \,(U+ Z \cN)^{-1} \,.
\end{equation}


\subsection{Prepotentials for Isometries of the c-map}
\label{section:prepotential}
Here we shall review the proof that isometries of $\M_{\rm h}$ whose Lie derivative on the $Sp(1)$-connection $\omega^x$ vanishes lead to prepotentials of the simple form \eqref{prepotential_no_compensator} \cite{Michelson:1996pn}. Let us assume that $k$ is an isometry of $\M_{\rm h}$ such that
\begin{equation} \label{Lie_der_omega}
 {\cal L}_k \omega^x \equiv \diff \omega^x (k,\cdot) + \diff (\omega^x(k)) = 0 \ .
\end{equation}
This implies that the Lie derivative of the $Sp(1)$ curvature two-forms $K^x$ \eqref{def_Sp(1)_curvature} vanishes
\begin{equation}
\begin{aligned}
 {\cal L}_k K^x =& \diff K^x (k,\cdot,\cdot) + \diff (K^x(k,\cdot)) \\ = & \tfrac12 \epsilon^{xyz} \diff(\omega^y\wedge \omega^z)(k,\cdot, \cdot) + \diff ( \diff \omega^x (k,\cdot) +  \tfrac12 \epsilon^{xyz} (\omega^y\wedge \omega^z)(k,\cdot) ) \\  = & \epsilon^{xyz} ( (\diff \omega^y\wedge \omega^z)(k,\cdot, \cdot) + \diff (\omega^y(k) \omega^z)) \\ = & \epsilon^{xyz} ( (\diff \omega^y(k,\cdot) \wedge \omega^z) + \diff (\omega^y(k)) \wedge \omega^z) = 0\ ,
\end{aligned}
\end{equation}
where we have also used \eqref{deriv_Sp(1)_curvature}. On the other hand, we can express the Lie derivative of $K^x$ via \eqref{deriv_Sp(1)_curvature}, \eqref{Pdef}, \eqref{Lie_der_omega} and \eqref{def_Sp(1)_curvature} as
\begin{equation}
\begin{aligned}
 {\cal L}_k K^x =& - \epsilon^{xyz} (\omega^y(k) K^z +\epsilon^{zx'y'} \omega^y\wedge  \omega^{x'} P^{y'}+\diff \omega^y P^z ) \\ =& - \epsilon^{xyz} (\diff \omega^z (\omega^y(k) - P^y)+ \epsilon^{zx'y'} (\tfrac12 \omega^y(k) \omega^{x'}\wedge \omega^{y'}+ \omega^y \wedge \omega^a P^b)) \\ = & - \epsilon^{xyz} (\omega^y(k) - P^y) K^z \ ,
\end{aligned}
\end{equation}
which can only vanish for $\omega^x(k) = P^x$, thus leading to \eqref{prepotential_no_compensator}.
One can check that the isometries given in \eqref{Killing} fulfil \eqref{Lie_der_omega} for the connection \eqref{quat_connection}.

\section{Stability of $\cN=1$ Vacua}\label{section:stability}

A vacuum which displays $\cN=2 \rightarrow \cN=1$ partial breaking supersymmetry should be stable. One can infer this by using a positive-energy theorem argument \cite{Cecotti:1984wn}. In this appendix, we shall present an alternative derivation of the same result by analysing the scalar potential $V$ and its derivatives.

We start from the Ward identity \eqref{potential_identity}, which we repeat here for convenience
\begin{equation} \label{potential_identity2}
 V \delta^{\cal A}_{\cal B} = -12 S_{\cal BC} \bar S^{\cal AC} + g_{i\bar \jmath} W^{i \cal {AC}} W^{\bar \jmath}_{\cal BC}
+ 2 N_\alpha^{\cal A} N_{\cal B}^\alpha \ .
\end{equation}
By contracting this with the product of unbroken generators $\epsilon^{\cal B}_1\epsilon^*_{1\,\cal A}$ and making use of \eqref{N=1conditions} we find the potential energy at the $\cN=1$ point, which is indeed non-positive and given by
 \begin{equation}\label{N=1pot}
V_{\cN=1} = -3 |\mu|^2 = \Lambda \ .
\end{equation}
Note that \eqref{potential_identity2} also states that for the broken
supersymmetry $\epsilon^{\cal B}_2$, the additional contributions of
$S_{\cal AB}$, $W^{i \mathcal{AM}}$ and $N_\alpha^{\cal A}$ have to
exactly cancel such that \eqref{N=1pot} holds.
Next we compute the derivatives of \eqref{potential_identity2}:
\begin{subequations} \label{potential_firstderivative}
\begin{eqnarray}\label{potential_firstderivative1}
\nabla_i V \delta^{\cal A}_{\cal B} &= & - 4 g_{i\bar \jmath }W^{\bar \jmath }_{\cal
  BC} \bar S^{\cal AC}  - W^{j \cal {AC}} {\cal M}_{\mathcal{BC}ij} +
N_\alpha^{\cal A} {\cal M}^{\alpha}_{i \mathcal{B}}
\ ,\\ \nonumber
\nabla_u V \delta^{\cal A}_{\cal B} &=& 6 \bar S^{\cal AC} N_{u({\cal BC})} + 6 N_u^{({\cal AC})} S_{\cal BC} - \tfrac{1}{2} {\cal
  M}^{( {\cal A}}_{\bar \jmath  \alpha} U^{\mathcal{C}) \alpha}_u W^{\bar \jmath }_{\cal BC} - \tfrac{1}{2} W^{i \cal {AC}} {\cal
  M}^{\alpha}_{i (\mathcal{B}} U_{\mathcal{C}) \alpha u} \\ &  & + ~ 8
   \bar S^{\cal AC} N_{u{\cal CB}} + 8 N_u^{\cal CA} S_{\cal BC}  + 2
  N_\alpha^{\cal A}  {\cal M}^{\alpha \beta}  U_{u\beta\mathcal{B}} + 2
  U_u^{\alpha \mathcal{A}} {\cal M}_{\alpha \beta}   N^\beta_{\cal B} \label{potential_firstderivative2}
  \ ,
\end{eqnarray}
\end{subequations}
where we have defined the sGoldstino matrix
\begin{equation} \label{sGoldstino-matrix}
 N_{u{\cal AB}}= U_{u {\cal A} \alpha} N^\alpha_{\cal B} \ .
\end{equation}
The matrices ${\cal M}_{\mathcal{AB}ij}$, ${\cal M}^{\alpha}_{i \mathcal{A}}$
and ${\cal M}_{\alpha \beta}$ are the mass matrices of the spin $1/2$
particles, which in the absence of gaugings of the vector multiplets are defined by~\cite{D'Auria:2001kv}
\begin{equation}\label{massmatrices}
\begin{aligned}
 {\cal M}_{\mathcal{AB}ik} \, = \, & g_{i \bar \jmath } \nabla_k W^{\bar \jmath }_{\cal AB}  \ , \\
 {\cal M}^{\alpha}_{i {\cal A}} \, = \, &  2 \nabla_i N^\alpha_{\cal A}  \ , \\
 {\cal M}^{\alpha \beta} \, = \,&  \tfrac12 U^{u{\cal A}\alpha} \nabla_u N_{\cal A}^\beta  \ .
\end{aligned}
\end{equation}
If we contract \eqref{potential_firstderivative} with $\epsilon^{\cal B}_1\epsilon^*_{1\,\cal A}$, it simplifies due to \eqref{N=1conditions} to give
\begin{equation} \label{potential_stationary}
(\nabla_i V)_{\cN=1} = 0 \ , \qquad (\nabla_u V)_{\cN=1} = 0 \ .
\end{equation}
Thus, the $\cN=1$ vacuum is a stationary point of the potential.

Next, we check the second derivatives. After contraction with $\epsilon^{\cal B}_1\epsilon^*_{1\,\cal A}$ and excessive use of \eqref{N=1conditions}, the result reads
\begin{equation} \label{potential_secondderivative}
\begin{aligned}
(\nabla_i \nabla_j V)_{\cN=1} =&
\bar\mu {\cal M}_{\mathcal{AB}ij} \epsilon^{\cal A}_1 \epsilon^{\cal B}_1/|\epsilon_1|^2
\ , \\
(\nabla_i \nabla_{\bar \jmath  } V)_{\cN=1} =&
-2 |\mu|^2 g_{i\bar \jmath } + g^{k\bar l}  ({\cal M}^\mathcal{AC}_{\bar \jmath  \bar l} \epsilon^*_{1\,\cal A}) ({\cal M}_{\mathcal{CB}ik} \epsilon^{\cal B}_1)/|\epsilon_1|^2  +\tfrac{1}{2} (\bar{\cal M}^\mathcal{A}_{\bar \jmath  \alpha} \epsilon^*_{1\,\cal A}) ({\cal M}^{\alpha}_{i \mathcal{B}} \epsilon^{\cal B}_1)/|\epsilon_1|^2   \ , \\
(\nabla_i \nabla_u V)_{\cN=1} =&
3 \bar\mu ({\cal M}^{\alpha}_{i (\mathcal{B}}\epsilon^{\cal B}_1) (U_{u \mathcal{A}) \alpha } \epsilon^{\cal A}_1)/|\epsilon_1|^2  + \tfrac{1}{2} ({\cal M}_{\mathcal{BC}ik} g^{k \bar \jmath }\epsilon^{\cal B}_1)   ({\cal M}^{(\mathcal{A}}_{\bar \jmath  \alpha}  U^{\mathcal{C}) \alpha}_u \epsilon^*_{1\,\cal A})/|\epsilon_1|^2 \\ & + {\cal M}_{i \alpha \mathcal{B}} \epsilon^{\cal B}_1 {\cal M}^{\alpha \beta} U_{u \mathcal{A} \beta } \epsilon_1^{\cal A}/|\epsilon_1|^2    \ , \\
(\nabla_u \nabla_v V)_{\cN=1} =&
- 6 (N_{u({\cal BC})}\epsilon^{\cal B}_1 ) (N_v^{({\cal AC})}\epsilon^*_{1\,\cal A} ) / |\epsilon_1|^2
+ 28 |\mu|^2  (U_{u \mathcal{A} \alpha} \epsilon_1^{\cal A}) (U_{v}^{\mathcal{B}\alpha} \epsilon^*_{1\,\cal B})/|\epsilon_1|^2
\\ & + 11  \bar\mu (U_{u \mathcal{A} \alpha} \epsilon^{\cal A}_1) {\cal M}^{\alpha \beta} (U_{v\mathcal{B}\beta} \epsilon^{\cal B}_1)  /|\epsilon_1|^2
+ 11  \mu (U_u^{\mathcal{A} \alpha} \epsilon^*_{1\,\cal A}) {\cal M}_{\alpha \beta} (U_v^{\mathcal{B}\beta} \epsilon^*_{1\,\cal B})  /|\epsilon_1|^2 \\ &
+ \tfrac{1}{2} ({\cal M}^{\alpha}_{i (\mathcal{C}}U_{\mathcal{B}) \alpha u}\epsilon_1^{\cal B})  g^{i\bar \jmath }({\cal M}^{(\mathcal{A}}_{\bar \jmath  \beta}U_{v}^{\mathcal{C}) \beta} \epsilon^*_{1\,\cal A})/|\epsilon_1|^2 \\ &
 + 4  (U_{u \mathcal{A} \beta} \epsilon^{\cal A}_1) {\cal M}^{\beta\alpha} {\cal M}_{\alpha\gamma} (U_{v}^{\mathcal{B}\gamma} \epsilon^*_{1\,\cal B}) /|\epsilon_1|^2
  \ .
\end{aligned}
\end{equation}
With these expressions we can identify the mass terms in the Lagrangian $\cal L$ to be
\begin{equation} \label{Mass_term}
\begin{aligned}
{\cal L}_{\rm mass} = & \tfrac12 t^i (\nabla_i \nabla_j V)_{\cN=1} t^j + t^i(\nabla_i \nabla_{\bar \jmath } V)_{\cN=1} \bar t^{\bar \jmath } + \tfrac12 \bar t^{\bar \imath} (\nabla_{\bar \imath} \nabla_{\bar \jmath } V)_{\cN=1} \bar t^{\bar \jmath } \\ & + t^i (\nabla_i \nabla_u V)_{\cN=1} q^u + \bar t^{\bar \imath } (\nabla_{\bar \imath } \nabla_u V)_{\cN=1} q^u + \tfrac12 q^u (\nabla_u \nabla_v V)_{\cN=1} q^v \\
= & -\tfrac94 |\mu|^2 t^i g_{i \bar \jmath } \bar t^{\bar \jmath } - \tfrac98 |\mu|^2 q^u h_{uv} q^v - 3 (N_{u({\cal BC})}\epsilon^{\cal B}_1 ) (N_v^{({\cal AC})}\epsilon^*_{1\,\cal A} ) / |\epsilon_1|^2 \\ &
+ \bar \Phi^{\cal A}_{\bar \jmath }(t,\bar t,q) g^{\bar \jmath i} \Phi_{i{\cal A}}(t,\bar t,q) + \tfrac12 \bar \Psi_\alpha(\bar t,q) \Psi^\alpha(t,q)  \ ,
\end{aligned}
\end{equation}
where we abbreviated
\begin{equation}
\begin{aligned}
\Phi_{k{\cal C}}(t,\bar t,q)=& ({\cal M}_{\mathcal{CB}ki}\epsilon^{\cal B}_1 t^i  + \tfrac12 \mu g_{k\bar \imath } \bar t^{\bar \imath }  \epsilon^*_{1\,\cal C} + \tfrac12 {\cal M}^{\alpha}_{k (\mathcal{C}}U_{\mathcal{B}) \alpha u}\epsilon_1^{\cal B} q^u )/|\epsilon_1| \ , \\
\Psi^\alpha(t,q)  = & ({\cal M}^{\alpha}_{i \mathcal{B}} \epsilon^{\cal B}_1 t^i + 2 {\cal M}^{\alpha\beta} U_{u \mathcal{B} \beta} \epsilon^{\cal B}_1 q^u + \tfrac{11}{2} \mu U_u^{\mathcal{B} \alpha} \epsilon^*_{1\,\cal B} q^u )/|\epsilon_1| \ .
\end{aligned}
\end{equation}

The first two terms in \eqref{Mass_term} give the Breitenlohner-Freedman bound in our conventions~\cite{Breitenlohner:1982jf}. Therefore, to ensure stability we have to show that the third (negative definite) term is compensated by the two last (positive definite) terms.
To do so, we contract \eqref{potential_firstderivative1} with $\epsilon^{\cal B}_1 t^i$ and \eqref{potential_firstderivative2} with $\epsilon^{\cal B}_1 q^u$ and add them together. After application of \eqref{potential_stationary} we find
\begin{equation} \label{stability_relation_1}
 6 \bar S^{\cal AC} N_{u({\cal CB})} \epsilon^{\cal B}_1 q^u /|\epsilon_1|=  W^{k \cal {AC}} \Phi_{k{\cal C}}(t,\bar t,q) -  N_\alpha^{\cal A}  \Psi^\alpha(t,q) \ ,
\end{equation}
and similarly
\begin{equation} \label{stability_relation_2}
 6 q^v \epsilon^*_{1\,\cal A}  N_v^{({\cal AC})} S_{\cal CB}/|\epsilon_1| =  \bar \Phi^{\cal C}_{\bar l}(t,\bar t,q) W^{\bar l}_{\cal CB} - \bar \Psi_\alpha(\bar t,q) N^\alpha_{\cal B} \ .
\end{equation}
Note that we added the further term $\tfrac12 \mu  W^{k \cal {AC}} g_{k\bar \imath } \bar t^{\bar \imath }  \epsilon^*_{1\,\cal C} = 0$.
If we multiply \eqref{stability_relation_1} with \eqref{stability_relation_2} and contract the free indices, we find
\begin{equation}\label{stability_relation}
\begin{aligned}
 0=&-36 q^v \epsilon^*_{1\,\cal A} N_v^{({\cal AC})} \bar S^{\cal CE} S_{\cal ED} N_{u({\cal DB})} \epsilon^{\cal B}_1 q^u / |\epsilon_1|^2 +  \bar \Phi^{\cal C}_{\bar l}(t,\bar t,q) W^{\bar l}_{\cal CB} W^{k \cal {BA}} \Phi_{k{\cal A}}(t,\bar t,q) \\ &
 +\bar \Psi_\beta(\bar t,q) N^\beta_{\cal A} N_\alpha^{\cal A}  \Psi^\alpha(t,q)
- \bar \Psi_\alpha(\bar t,q) N^\alpha_{\cal A} W^{k \cal {AC}} \Phi_{k{\cal C}}(t,\bar t,q) -  \bar \Phi^{\cal C}_{\bar l}(t,\bar t,q) W^{\bar l}_{\cal CA} N_\alpha^{\cal A}  \Psi^\alpha(t,q) \ .
\end{aligned}
\end{equation}
By using the completeness condition for the supersymmetry generators
\begin{equation}\label{completeness_condition}
\frac{\epsilon^*_{1\,\cal A} \epsilon_1^{\cal B}}{|\epsilon_1|^2} + \frac{\epsilon^*_{2\,\cal A} \epsilon_2^{\cal B}}{|\epsilon_2|^2} = \delta_{\cal A}^{\cal B}  \ ,
\end{equation}
together with \eqref{sGoldstino-matrix} and \eqref{N=1conditions} we find
\begin{equation} \label{simplification_S}
 \bar S^{\cal EC} S_{\cal ED} (N_{u({\cal BC})} \epsilon^{\cal B}_1) (N_v^{({\cal AD})} \epsilon^*_{1\,\cal A}) = \tfrac{|S\cdot \epsilon_2|^2}{|\epsilon_2|^2}  (N_{u({\cal BC})} \epsilon^{\cal B}_1) (N_v^{({\cal AC})} \epsilon^*_{1\,\cal A}) \ .
\end{equation}
With the help of \eqref{simplification_S}
we derive from \eqref{stability_relation} the relation
\begin{equation} \label{equality2}
\begin{aligned}
 - 3 (N_{u({\cal BC})}\epsilon^{\cal B}_1 ) (N_v^{({\cal AC})}\epsilon^*_{1\,\cal A} )/|\epsilon_1|^2 = &
 - \tfrac{1}{12} \tfrac{|\epsilon_2|^2}{|S\cdot \epsilon_2|^2}  \bar \Phi^{\cal B}_{\bar \jmath }(t,\bar t,q) W^{\bar \jmath}_{\cal CB} W^{i{\cal AC}} \Phi_{i{\cal A}}(t,\bar t,q) \\ & - \tfrac{1}{12} \tfrac{|\epsilon_2|^2 }{|S\cdot \epsilon_2|^2}   \bar \Psi_\beta(\bar t,q) N^\beta_{\cal A} N_\alpha^{\cal A} \Psi^\alpha(t,q) \\ &
 + \tfrac{|\epsilon_2|^2 }{12|S\cdot \epsilon_2|^2}\bar \Psi_\alpha(\bar t,q) N^\alpha_{\cal A} W^{k \cal {AC}} \Phi_{k{\cal C}}(t,\bar t,q)\\ & + \tfrac{|\epsilon_2|^2 }{12|S\cdot \epsilon_2|^2} \bar \Phi^{\cal C}_{\bar l}(t,\bar t,q) W^{\bar l}_{\cal CA} N_\alpha^{\cal A}  \Psi^\alpha(t,q)\ .
\end{aligned}
\end{equation}
From \eqref{potential_identity2}, we find the relation
\begin{equation} \label{SWN_identity}
 |S\cdot \epsilon_2|^2 = \tfrac14|\mu|^2|\epsilon_2|^2 + \tfrac{1}{12} |\epsilon_2|^2  |W|^2 + \tfrac{1}{6} |\epsilon_2|^2  |N|^2 \ .
\end{equation}
Thus, starting from \eqref{potential_secondderivative}, we find with \eqref{equality2} and \eqref{SWN_identity} the following result for the scalar mass terms in the Lagrangian
\begin{equation}\label{massLagrangian}
\begin{aligned}
{\cal L}_{\rm mass} = &
 -\tfrac94 |\mu|^2 t^i g_{i \bar \jmath } \bar t^{\bar \jmath } - \tfrac98 |\mu|^2 q^u h_{uv} q^v \\ & + \tfrac{|\epsilon_2|^2}{12 |S\cdot \epsilon_2|^2} \bar \Psi_\beta(\bar t,q) \left( (|N|^2 + \tfrac32 |\mu|^2) \delta^\beta_\alpha - N^\beta_{\cal A} N_\alpha^{\cal A} \right)  \Psi^\alpha(t,q)
 \\ & + \tfrac{|\epsilon_2|^2}{12  |S\cdot \epsilon_2|^2} \bar \Phi^{\cal B}_{\bar \jmath }(t,\bar t,q) \left((|W|^2+ 3|\mu|^2)g^{\bar \jmath i} \delta^{\cal A}_{\cal B} - W^{i {\cal AC}} W^{\bar \jmath}_{\cal CB} \right) \Phi_{i{\cal A}}(t,\bar t,q) \hfill \\ & + \tfrac{|\epsilon_2|^2 }{12 |S\cdot \epsilon_2|^2} (\bar \Phi^{\cal A}_{\bar \jmath }(t,\bar t,q) N_\alpha^{\cal B} + \bar \Psi_\alpha(\bar t,q)W^{k \cal {AB}}  g_{k\bar \jmath }) \\ & \hskip1.2cm \cdot g^{\bar \jmath  i}(N^\alpha_{\cal B} \Phi_{i{\cal A}}(t,\bar t,q) + g_{i\bar l}W^{\bar l}_{\cal BA} \Psi^\alpha(t,q)) \ .
\end{aligned}
\end{equation}
By again using the completeness relation \eqref{completeness_condition}, we find
\begin{equation}
 W^{i \cal {AC}} W^{\bar \jmath }_{\cal CB} = \epsilon^*_{2\,\cal A} \epsilon_2^{\cal B} W^{i \cal {CD}} W^{\bar \jmath }_{\cal CD}/|\epsilon_2|^2 \le \delta_{\cal A}^{\cal B} W^{i \cal {CD}} W^{\bar \jmath }_{\cal CD}\ .
\end{equation}
Furthermore, since $W^{i \cal {AB}} W^{\bar \jmath }_{\cal AB}$ and $N_\alpha^{\cal A} N^\beta_{\cal A}$ are hermitian matrices, we can establish the following inequalities
\begin{equation} \label{inequality}
\begin{aligned}
 W^{i \cal {AC}} W^{\bar \jmath }_{\cal CB} & \le \delta_{\cal A}^{\cal B} g^{i \bar \jmath } |W|^2 \ , \\
 N_\alpha^{\cal A} N^\beta_{\cal A} & \le \delta_\alpha^\beta |N|^2 \ .
\end{aligned}
\end{equation}
This shows that \eqref{massLagrangian} is bounded from below by the Breitenlohner-Freedman bound \cite{Breitenlohner:1982jf} and we have a stable minimum.

\bibliographystyle{JHEP-2}

\providecommand{\href}[2]{#2}\begingroup\raggedright\endgroup

\end{document}